%% file: manuscript-arxiv.tex
\documentclass[final,5p,twocolumn]{elsarticle}

\input{preamble.tex}

\journal{arXiv}

\begin{document}

\begin{frontmatter}

\title{Thermal performance estimation for cryogenic storage tanks:
  Application to liquid hydrogen}

\affiliation[SINTEF]{
  organization={SINTEF Energy Research},
  city={Trondheim},
  country={Norway}}
\affiliation[NTNU]{
  organization={Department of Mathematical Sciences, Norwegian University of Science and Technology},
  city={Trondheim},
  country={Norway}}
\author[SINTEF]{Ailo Aasen}
\ead{ailo.aasen@sintef.no}
\author[SINTEF,NTNU]{Sindre Stenen Blakseth}
\author[NTNU]{André Massing}
\author[SINTEF]{Petter Nekså}
\author[SINTEF]{Magnus Aa.\ Gjennestad}

\date{\today}

\begin{abstract}
  The design of cryogenic liquid storage solutions requires accurate
  methods for estimating heat ingress, from the material level to the
  tank level. For insulation materials, thermal performance is usually
  measured using ambient conditions and liquid nitrogen at
  \SI{77}{\kelvin} as boundary temperatures. A key question is
  how much heat ingress increases when storing liquid hydrogen (\lhto)
  at \SI{20}{\kelvin}. We derive theoretical bounds on the increased
  heat ingress, and show that it remains below
  \SI{26}{\percent}. Additionally, we demonstrate that heat ingress is
  much more sensitive to the warm boundary temperature than the cold
  boundary temperature. At the tank level, we compare two methods for
  assessing the steady-state thermal performance of cryogenic tanks:
  thermal network models and the heat equation solved with the finite
  element method. The latter offers high accuracy and adaptability for
  complex geometries, while thermal network models benefit from
  simplicity, speed and robustness. We apply both approaches to a
  self-supported \lhto tank concept for maritime transport and analyze
  sensitivity to structural support thickness, warm boundary
  temperature, and choice of insulation material. The
  thermal network model can estimate heat ingress with $\sim
  \SI{1}{\percent}$ error and the cold-spot temperature with error
  less than \SI{1}{\kelvin}.
\end{abstract}


\end{frontmatter}

\section{Introduction}
\label{sec:introduction}

A major hurdle for transitioning to renewable energy sources is that
these sources do not necessarily provide energy exactly when or where
it is needed. This challenge can be addressed through large-scale
production, transport and storage of clean energy carriers, such as
hydrogen (\ce{H2}) \cite{stolten2016hydrogen,Yin2024,Ma2024,Muthukumar2023,karayel2023comprehensive}.

Hydrogen is an energy carrier that can serve as a clean and
renewable fuel, but a hydrogen economy depends on effective storage
and transport mechanisms. Since hydrogen gas at ambient conditions has
a low density of \SI{0.09}{\kilo\gram\per\meter\cubed}, storing it in
this state requires huge volumes. If local geological conditions
permit, large-scale storage of gaseous hydrogen in salt caverns may be
a cost-effective option~\cite{Andersson2019,Ma2024}. Another option is to increase density
prior to storage. Strategies include chemically binding the hydrogen
in liquid organic hydrogen carriers~\cite{preuster2017liquid,zheng2021current},
physisorption on metal-organic frameworks (MOFs)~\cite{park202220},
compression and
liquefaction~\cite{krasae2010development,al2022hydrogen}.

Liquefied hydrogen (\lhto) has a density at ambient pressure of
\SI{71}{\kilo\gram\per\meter\cubed}, roughly three orders of magnitude
higher than the gas phase (\Cref{tab:phys_props}). \lhto has
exceptionally high purity, which after regasification enables direct
use in high-purity applications such as fuel cells \cite{stolten2016hydrogen,ustolin2022extensive}. Since the \ce{H2}
is not chemically bonded to heavy components (e.g.\ in \ce{NH3}) or
adsorbed in other materials (e.g.\ MOFs), \lhto has an unusually high
gravimetric energy density and can be converted directly to pure
\ce{H2} gas without further processing. Since \lhto can exist at
ambient pressure, storage tanks can also be sized thinner than
pressurized containers, reducing cost. An important drawback is
hydrogen's extremely low normal boiling point of \SI{-253}{\celsius}, and the
large associated energy input required for the liquefaction
process. With today's liquefaction technology, a third of the energy
content of gaseous hydrogen is needed to liquefy
it~\cite{berstad2009comparison}. However, for large-scale optimized
plants, it is predicted that this may be reduced to below
\SI{20}{\percent}~\cite{Cardella2017b,Wilhelmsen2018}. Moreover, some exergy invested
into liquefaction can be reclaimed during regasification.

\begin{table*}[tbp]
  \caption{Thermophysical properties of liquid hydrogen, liquid
    methane and liquid nitrogen at their respective normal boiling
    points ($p=\SI{1}{\atm}$). Here $T$ is temperature, $\rho$ is
    density, $\Delta H_\vap$ is enthalpy of vaporization per liquid
    mass or volume, $c_p$ is isobaric heat capacity, $k$ is thermal
    conductivity, $\mu$ is viscosity, and $\gamma$ is surface
    tension~\cite{Leachman2009,Span2000Nitrogen,Setzmann1991Methane,bell2014pure}.
    \label{tab:phys_props}} \centering
  \begin{tabular}{lcccccccc}
  \toprule
                  &             $T$ &          $\rho$ & $\Delta H_\vap$ & $\Delta H_\vap$ &           $c_p$ &             $k$ &           $\mu$ &        $\gamma$ \\
                  &               K &   \si{\kg/\m^3} &    \si{\kJ/\kg} &   \si{\mega\J/\m^3} & \si{\kJ/\kg \K} &    \si{\W/\K\m} & \si{\micro\Pa\s} & \si{\milli\N/\m} \\
  \midrule
         Hydrogen &            20.4 &            70.8 &             449 &            31.8 &            9.77 &           0.104 &            13.5 &            1.91 \\
         Nitrogen &            77.4 &             806 &             199 &             161 &            2.04 &           0.145 &             161 &            8.88 \\
          Methane &             112 &             422 &             511 &             216 &            3.48 &           0.184 &             117 &            12.9 \\
  \bottomrule
  \end{tabular}
  \end{table*}

To make an \lhto market economically feasible it must be scaled up by
orders of magnitude, and a crucial technology
gap is the development of cost-effective large-scale storage tanks~\cite{Ratnakar2021}.
\lhto must be stored in tanks with sophisticated insulation to limit
boil-off, with tank design depending on the application~\cite{Adler2024b}. The
stationary tanks, and the mobile tanks on ships, aircraft, rockets,
trains and trucks all feature different design problems related to
loading/offloading, venting, optimal shape, sloshing and risks in
general. Use of smaller stationary tanks is today routine, although
tanks for large-scale \lhto storage and transport similar to those
used for liquefied natural gas (LNG), i.e.\ $\gtrsim \SI{40000}{\meter\cubed}$, have not yet been
built. One challenge stems from the thermophysical peculiarities of
\lhto, which are elucidated when compared to liquid methane and liquid
nitrogen (\Cref{tab:phys_props}). Its low boiling point temperature
and low volumetric latent heat necessitate high thermal insulation
performance~\cite{Ratnakar2021}.

Recently, there has been a surge of interest in the thermal modeling of
liquid hydrogen tanks. Several studies focus on lumped-phase thermodynamic models
for studying the pressurization rate and thermal
stratification~\cite{joseph2017effect,al2022modelling,majumdar2008numerical,Barsi2008,barsi2013investigation,Adler2024b,Matveev2023,mendez2021enabling}.
Joseph et al.~\cite{joseph2017effect} observe that both effects
increase with decreasing insulation thickness, but a crucial unknown is the rate of heat transport across the interface \cite{al2022modelling,Adler2024b,Matveev2023}, which in
practice is treated as a fudge factor tuned to experiments. Other studies focus
more on the
insulation~\cite{ratnakar2023effective,Babac2009,zheng2019thermodynamic,Zheng2019,Yin2024,Jiang2021,liu2016thermal,Hofmann2006,Wang2020},
where the application of multi-layer insulation (MLI) and vapor-cooled
shields are recurring topics. The general conclusion is that both MLI and
vapor-cooled shields have a large potential to reduce boil-off, but
their economic and technical feasibility for large-scale tanks is to
date unresolved.

Yet other studies focus on the heat flow and temperatures in the
insulation and tank, which are important issues to consider when
analyzing thermal performance and thermal stresses. The work by
Tapeinos et
al.~\cite{tapeinos2016design,tapeinos2019evaluationI,tapeinos2019evaluationII}
combined experiments with finite element analysis of thermal and
structural aspects for a novel \lhto multi-sphere tank configuration.
Johnson et al.~\cite{johnson2022analysis} studied heat transfer around
localized hot and cold spots. Such spots may lead to increased thermal
stresses while also affecting \lhto flow patterns and heat transfer
from the ambient. A recent work by Mantzaroudis and
Theotokoglou~\cite{Mantzaroudis2023} emphasizes that accounting for
the temperature-dependence of thermophysical properties is crucial.

When optimizing an \lhto tank design, models that are valid over a wide space of
design parameters are needed. A notable recent work by Adler and
Martins~\cite{Adler2024b} presents an open-source tank model for liquid hydrogen
storage which can be coupled with gradient-based optimization to find, e.g., the
minimum weight for given storage restraints.

Use and transport of \lhto in shipping is an emerging field receiving
significant attention
~\cite{Abe1998,Scurlock2017,Ratnakar2021,alkhaledi2022hydrogen,Cristea2020}.
Already in 1998, Abe presented a conceptual design of a
\SI{200000}{\meter\cubed} hydrogen tanker based on LNG technology. The use of
\lhto as maritime fuel holds the promise to eliminate harmful emissions from
shipping, which corresponds to roughly 3\% of global greenhouse gas emissions
and more than 12\% of SOx and NOx emissions~\cite{Wang2020}. Current bottlenecks
include cost and lack of regulations, infrastructure, and qualification of materials \cite{ustolin2022extensive,Wang2020}.
Ratnakar et al.~\cite{Ratnakar2021} emphasized the possible advantages of
\ce{LH2} storage and transport due to the similarity with LNG technology and
existing \ce{LH2} technology in space programs.
However, they noted that ``for commercial widespread use and
feasibility of hydrogen technology, it is of utmost importance to
develop cost-effective and safe technologies for storage and
transportation of \ce{LH2} for use in stationary applications as well
as offshore transportation.'' They pointed to the need for
reduced-order models that can easily analyze heat transport across a
wide parameter space. Furthermore, they noted that while there is a
significant amount of experimental data on the thermal performance of
insulation materials in the literature, much of this is based on
measurements with \ce{LN2} and a cold boundary temperature (CBT) of
\SI{77}{\kelvin}. It is therefore some uncertainty as to how these
materials will perform in \ce{LH2} applications with cold boundary
temperatures down towards \SI{20}{\kelvin}.

The present work aims to address some of these
issues. First, in \Cref{sec:theory}, we briefly reiterate some of the
formalism of heat transport in insulation and solid materials and show
that this immediately leads to some new insights when applied to \lhto
storage. In particular, we present a procedure to quantify the
uncertainty in heat ingress that is a result of using models and
measurements obtained with \ce{LN2} and a CBT of \SI{77}{\kelvin}.
Next, we present a thermal network model, a type of reduced-order
model, for cryogenic tanks that can account for cold spots through a
correction. The numerical methods used are discussed in
\Cref{sec:numerical}.

In \Cref{sec:results}, we test the presented procedure to quantify the
uncertainty in heat ingress, apply it to several measurements and
models available in the literature and discuss the results. We also
argue that heat transport may be particularly sensitive to changes in
the warm boundary temperature (WBT), and that cold spots induced by
support structures acting as thermal bridges may therefore have an
important impact on the total heat ingress.

In \Cref{sec:case-study}, we perform a case study where we apply the
network model to a self-supported, non-pressurized tank intended for
ship transport of \ce{\lhto}. We compare the network model results to
full solutions of the heat equation obtained with the finite element
method (FEM). Finally, a sensitivity analysis on the selected case is
performed and the impact of structural support size, ambient
temperature and choice of insulation material is
quantified. Furthermore, the uncertainty in total heat ingress and
boil-off rate resulting from use of models and measurements based on
experiments with \ce{LN2} is presented and discussed.

\section{Heat transfer modeling}
\label{sec:theory}

In a piece of material placed between two heat baths at differing
temperatures $\Tc < \Tw$, a heat flux $\vec{q}$ will develop.
For an insulation material, e.g.\ a porous bulk-fill material,
this can be decomposed into contributions from gas conduction, solid
conduction, gas convection and radiation,
\begin{equation}
  \label{eq:q}
  \vec{q} = \vec{q}_\cond^\gas + \vec{q}_\cond^\solid + \vec{q}_\conv
  + \vec{q}_\rad.
\end{equation}
Detailed modeling of each heat transfer mode can be complex and
simpler models often offer sufficient accuracy. One common approach is
to aggregate all modes into an equivalent conduction mode, having an
apparent thermal conductivity $k\pp{T}$ \textit{defined} through
Fourier's law
\begin{equation}
  \vec{q} = - k \pp{T} \grad T. \label{eq:fouriers-law}
\end{equation}
For a porous material the thermal conductivity also depends on the
pressure $p$ of the interstitial gas, i.e.\ $k \pp{T,p}$. Unlike the
temperature, however, the pressure $p$ will generally be constant
throughout the material. We can therefore write $k \pp{T}$, with the
understanding that this function is valid for a particular gas
pressure. For a solid material, \Cref{eq:fouriers-law} can of course
be used directly with $k$ describing the thermal conductivity of the
material.

In steady-state, local energy balance is described by $\div \vec{q} =
0$. Inserting Fourier's law~\eqref{eq:fouriers-law} the energy balance
equation becomes the steady-state heat equation,
\begin{equation}
  \div \pp{ -k\pp{T} \grad T} = 0. \label{eq:heat-eqn}
\end{equation}
For \Cref{eq:heat-eqn} to have a unique solution, boundary conditions
must be specified. For example, one may specify the boundary
temperature (Dirichlet condition) or a temperature-independent heat
flux (Neumann condition). Another option is a Robin condition, where
the heat flux $q$ normal to the boundary is determined by some
convective heat transfer coefficient $\htc$ and ambient temperature
$T_\amb$, so that
\begin{equation}
  q = -k \pp{T} \grad T \vec{\cdot} \vec{n} = \htc \pp{T - T_\amb}.
  \label{eq:htc_bc}
\end{equation}
Here, $\vec{n}$ is the outward unit normal vector of the boundary.

\subsection{Material properties}
\label{sec:material_properties}

The apparent thermal conductivity of both insulation materials and structural
materials can vary significantly between typical ambient temperatures and
typical \lhto storage temperatures~\cite{barron2017cryogenic}, and we next consider models for this
temperature dependence.

Hofmann~\cite{Hofmann2006} presented the following largely empirical
model
\begin{align}
  \label{eq:k_hofmann}
  k \pp{T} &= a + bT^c,
\end{align}
were $a$, $b$ and $c$ are parameters that need to be fitted to
experimental results, and which implicitly are functions of the gas
pressure. Hofmann fitted parameters for seven insulations: nitrogen
gas at atmospheric pressure, evacuated perlite, evacuated microglass
spheres, evacuated fibreglass, an evacuated multilayer insulation
system, and a hypothetical insulation of ideally arranged
reflectors in perfect vacuum.

NIST~\cite{NISTCryogenicsWebpage} has collected data on thermal
conductivities of a range of structural and insulation materials and
give empirical models on the form
\begin{align}
  \log_{10} \pp{k \pp{T}} &= \sum_{i=0}^{N} c_i \pp{\log_{10} \pp{T}}^i,
  \label{eq:k_NIST}
\end{align}
where $c_i$ are fitted parameters.

Ratnakar et al.~\cite{ratnakar2023effective} reviewed thermal
conductivity modeling in cryogenic insulation materials, and
introduced a first-principles-based functional form
\begin{align}
  \label{eq:k_ratnakar}
  k \pp{T, p} &= AT^m + BT^3 + \frac{CT^{E}}{1 + DT/p}.
\end{align}
Here $m$, $A$, $B$, $C$, $D$ and $E$ are non-negative parameters,
which are independent of temperature and pressure. The first two terms
account for solid conduction and radiation, respectively. The third
term accounts for pressure-dependent gas conduction. In writing the
conductivity as a sum of conductivities from different heat transfer
modes, one implicitly assumes that the thermal resistances act in
parallel with negligible coupling effects between the heat transfer
modes.

The third term in \Cref{eq:k_ratnakar} is the key innovation over
Hofmann's correlation, as it incorporates the pressure-dependence
explicitly. The functional form is inspired by kinetic gas theory, and
corresponds to a gas conductivity $k^{\rm gas} \pp{T,p} =
k_{\rm c} \pp{T}/\pp{1+\Kn}$, where $k_{\rm c} \pp{T}=C T^E$ is the
pressure-independent continuum limit conductivity and $\Kn=DT/p$ is
the Knudsen number. $\Kn$ is the gas mean free path divided by the
length scale of voids in the material. It is usually close to zero at
atmospheric pressure, so that $k^{\rm gas}\pp{T, \SI{1}{atm}} \approx
k_{\rm c} \pp{T}$. Since $k_{\rm c} \pp{T}$ is usually known,
e.g.\ for helium, hydrogen and components of air, the temperature
dependence $E$ is taken as known. The prefactor $C$ cannot be
similarly deduced since it is depends on the porosity of the material,
with more voids implying more gas conduction heat transfer and thus a
larger $C$.

\subsection{Heat flow in one-dimensional geometries}
One is often interested in steady-state heat flow through simple,
effectively one-dimensional geometries between boundary temperatures
$\Tc$ and $\Tw$, e.g.\ normally through a planar slab or radially in a
cylindrical or spherical geometry. Fourier's
law~\eqref{eq:fouriers-law} can then be integrated directly, yielding
the well-known factorization of the heat flow $Q$
as~\cite{barron2017cryogenic}
\begin{align}
  \label{eq:Q_from_K}
  Q &= S \, K \pp{\Tc, \Tw},
\end{align}
where the thermal conductivity integral,
\begin{align}
  K &= \int_{\Tc}^{\Tw} k\pp{T} \dd T, \label{eq:K-def}
\end{align}
depends only on boundary temperatures and material properties, and the
shape factor
\begin{align}
  S &= \qty(\int_{\xc}^{\xh} \frac{1}{A\pp{x}} \ \dd x)^{-1}, \label{eq:S-def}
\end{align}
depends only on the geometry. Calculation of $S$ requires an integral
involving the area $A\pp{x}$ between the positions $\xc$ and $\xh$ of
the cold and warm boundaries. The quantity $K$, on the other hand, is
a \textit{geometry-independent} quantity that can be used to compare
the thermal performance of materials. It follows from
\Cref{eq:Q_from_K} that the SI value of $K$ equals the heat flux (in
\si{\watt\per\meter^2}) through a \SI{1}{\meter} thick slab.

The thermal conductivity integral can also be expressed by the
\textit{effective thermal conductivity},
\begin{equation}
  \keff \pp{\Tc,\Tw} = \frac{K\pp{\Tc,\Tw}}{\Tw-\Tc}, \label{eq:keff}
\end{equation}
i.e.\ the average thermal conductivity over a temperature
interval. For consistency one also defines
$\keff(T,T)=k(T)$.  The quantities $\keff \pp{\Tc,\Tw}$ and $K \pp{\Tc,\Tw}$ are
equivalent in the sense that they both enable calculating the heat
flow through a slab of material with boundary temperatures $\Tc$ and
$\Tw$. Using the effective thermal conductivity, one may rewrite
\Cref{eq:Q_from_K} as
\begin{align}
  \label{eq:Q_from_R}
  Q =\Delta T / R
\end{align}
where $\Delta T = \Tw-\Tc$ and the thermal resistance is
\begin{align}
  \label{eq:R_from_k_eff}
  R = 1 /(S k_\eff).
\end{align}

We stress that the heat flow factorization in \Cref{eq:Q_from_K} is
derived assuming the validity of Fourier's law~\eqref{eq:fouriers-law}.
That Fourier's law is valid even for a
composite material having several modes of heat transfer leads to a
significant simplification of heat flow calculations since local heat
flow then depends only on local temperature and temperature
gradients. In addition to simplifying thermal performance analysis,
\Cref{eq:Q_from_K} also leads to some important insights for
deep cryogenic applications, as we discuss next. To our
knowledge these points have not previously been made in heat transfer
textbooks~\cite{barron2017cryogenic,incropera_ed6,Lienhard2019}
or in the research literature.

\subsection{Concavity Hypothesis -- Extrapolating measured thermal performance to {\lhto} temperatures}
\label{sec:concavity}

Thermal performance measurements for cryogenic insulation materials
are often made using experimental setups with CBT provided by liquid
nitrogen (\ce{LN2}) at its normal boiling point of
\SI{77}{\kelvin}~\cite{Ratnakar2021}. Using \ce{LN2} for this purpose
is convenient and repeating measurements with e.g.\ \lhto is more
challenging and costly. Consequently, there is limited material
property data available for thermal modeling of systems at or close to
the boiling point temperature of \lhto (\SI{20}{\kelvin}). This
section offers a partial remedy. We shall consider what can still be
inferred, with reasonable confidence, in spite of missing data for
heat transfer with low CBTs. This will be done by giving upper and
lower bounds for the heat ingress obtained when the CBT is reduced
below temperatures for which one has reliable information.

We consider three temperatures $\Tc < \Tm < \Tw$. Here, $\Tw$ now
represents a typical ambient temperature (e.g.\ \SI{293}{\kelvin}),
$\Tc$ represents some low temperature (e.g.\ \SI{20}{\kelvin}), and
$\Tm$ some intermediate temperature (e.g.\ \SI{77}{\kelvin}). We may
assume that there is an underlying function describing the
apparent thermal conductivity of the insulation material, for which 
have some partial information. In the first case we only know the integral $K\pp{\Tm,\Tw}$,
obtained from measuring the heat flow through the material at a fixed
CBT $\Tm$ (e.g.\ using an \ce{LN2} cryostat). Another case is
that we also know the underlying function $k\pp{T}$, but only for temperatures down
to $\Tm$. We consider both cases here and consider what can still be inferred
about the thermal performance in spite of missing information for
temperatures below $\Tm$.

\begin{figure}[htbp]
  \centering
  \includegraphics[width=0.9\columnwidth]{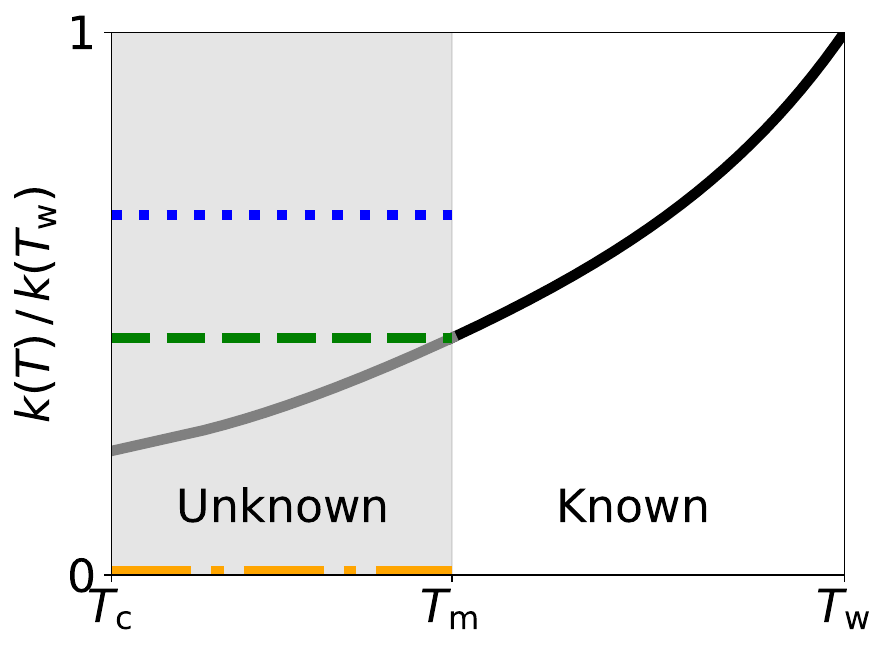}
  \caption{Illustration of normalized thermal conductivity $k \pp{T}$
    (black curve) and methods of extrapolating to low temperatures
    (colored curves). In the region between $\Tm$ and $\Tw$ (marked
    ``known''), we know either the underlying function $k\pp{T}$ or
    its integral $K \pp{\Tm,\Tw}$. In the region between $\Tc$ and
    $\Tm$ (marked ``unknown''), both are unknown. The blue dotted curve
    is the average value over the known region, the green dashed curve
    is the constant $k \pp{\Tm}$, and the orange dash-dotted curve is
    zero.}
  \label{fig:k-extrapolation-strategies}
\end{figure}

From \Cref{eq:Q_from_K}, the heat ingress is determined by
$K\pp{\Tc,\Tw}$. It follows from \crefrange{eq:K-def}{eq:keff} that
\begin{equation} \label{eq:K-decomposition}
  K\pp{\Tc, \Tw} = K\pp{\Tm, \Tw} + \keff\pp{\Tc, \Tm} \pp{\Tm - \Tc}.
\end{equation}
Herein, $K\pp{\Tm, \Tw}$ now represents the thermal conductivity
integral measured in cryostat experiments or the integral of the model
$k\pp{T}$ over its range of validity. To make further progress one
must estimate $\keff\pp{\Tc, \Tm}$. We shall do this by supplying
upper and lower bounds.

A lower bound follows from the fact that $k\pp{T}$ is positive, and
thus
\begin{equation}
  \keff(\Tc,\Tm) \ge 0.
\end{equation}
This lower bound is illustrated as the dash-dotted orange line in
\Cref{fig:k-extrapolation-strategies} in relation to a hypothetical
underlying function $k\pp{T}$. In terms of the thermal conductivity
integral this is equivalent to
\begin{equation}
   K \pp{\Tc, \Tw} \ge K_{\rm min} = K\pp{\Tm, \Tw}.
\end{equation}
The lower bound expresses that lowering the CBT will increase heat
ingress.

To estimate an upper bound, we assume that the unknown underlying
thermal conductivity $k\pp{T}$ is a monotonically increasing function
of temperature. This implies that $\keff \pp{\Tc,\Tm} \le \keff
\pp{\Tm,T_1}$ whenever $\Tm \le T_1$, and hence any choice of $T_1$
will yield an upper bound. Choosing $T_1=\Tw$ yields
\begin{equation}
  \keff \pp{\Tc,\Tm} \le \keff \pp{\Tm,\Tw} \label{eq:loose_bound}.
\end{equation}
This upper bound is illustrated as the dotted blue line in
\Cref{fig:k-extrapolation-strategies}. By inserting this bound into
\Cref{eq:K-decomposition}, we thus obtain
\begin{equation}
\label{eq:loose_bound_K}
\begin{aligned}
  K \pp{\Tc,\Tw} &\le K_{\max}^{\rm int}, \\ K_{\max}^{\rm int} &=
  K\pp{\Tm,\Tw} \frac{\Tw-\Tc}{\Tw-\Tm}.
\end{aligned}
\end{equation}
We emphasize that $K_{\max}^{\rm int}$ may be obtained by a single heat
flow measurement between $\Tm$ and $\Tw$ and may therefore be used
when cryostat measurements of heat flow between these two temperatures
are available. The superscript int alludes to it being proportional to
the integral $K \pp{\Tm,\Tw} = \int_{\Tm}^{\Tw} k\pp{T} \dd T$.

If one has measured $\keff \pp{\Tm, T_1}$ for decreasingly warm
boundary temperatures $T_1$, one obtains increasingly tight upper
bounds. In the event that a model for $k\pp{T}$ is known and valid
down to a temperature $\Tm$, one obtains the tightest bound when
$T_1=\Tm$, namely
\begin{equation}
  \keff \pp{\Tc, \Tm} \le \keff \pp{\Tm,\Tm} = k \pp{\Tm}.
\end{equation}
This upper bound is illustrated as the dashed green line in
\Cref{fig:k-extrapolation-strategies}. The corresponding upper bound
on the thermal conductivity integral is given by
\begin{equation}
  \label{eq:tight_bound}
  \begin{aligned}
    K\pp{\Tc,\Tm} &\le K_{\max}^{\rm diff}, \\ K_{\max}^{\rm diff} &= K
    \pp{\Tm,\Tw} + k\pp{\Tm} \pp{\Tm - \Tc}.
  \end{aligned}
\end{equation}
The superscript diff indicates that it depends on differential
information, as $k \pp{\Tm} = - \pp{ \partial K(\Tm, \Tc)/\partial
  \Tm}_{\Tc}$.
  
We now have that $K_{\min} \le K \le K_{\max}^{\rm diff} \le K_{\max}^{\rm int}$.
By multiplying with the shape factor $S$, we
get for the heat ingress $Q=SK$ that
\begin{align}
  \label{eq:concavity-Q}
  &Q_{\min} \le Q \le Q_{\max}^{\rm diff} \le Q_{\max}^{\rm int}, \quad \text{where} \\[3pt]
  Q_{\min} &= Q \pp{\Tm,\Tw}, \\
  Q_{\max}^{\rm diff} &= Q\pp{\Tm,\Tw} - \pp{\frac{\partial Q\pp{\Tm,\Tw}}{\partial \Tm}}_{\Tw} (\Tm-\Tc), \label{eq:Qmaxdiff} \\
  Q_{\max}^{\rm int} &= Q\pp{\Tm,\Tw} \frac{\Tw-\Tc}{\Tw-\Tm}. \label{eq:Qmaxint}
\end{align}
An equivalent statement is that \textit{heat flow is a decreasing,
  concave function of the cold boundary temperature.} We
therefore refer to \eqref{eq:concavity-Q} as the Concavity
Hypothesis. As it makes no reference to the apparent thermal conductivity $k$, it
can also be applied to materials where it is unclear how to define $k$ (e.g.\ MLI).

\subsection{Network model}
\label{sec:network-model}
One way to estimate the heat ingress in more complex geometries, such
as tanks, is to use a thermal network model. This is a
well-established
approach~\cite{majumdar2008numerical,ogunsola2015application} that
makes some simplifying assumptions about the geometry and the
available paths for heat flow. Here, we employ a version that accounts
fully for the temperature dependence of the thermal conductivity.

The thermal network is made up of nodes, each with an index $i$ and a
temperature $T_i$. The nodes are connected by thermal resistances
$R_{ij}$ such that the heat flow $Q_{ij}$ from node $i$ to node $j$ is
\begin{equation}
  Q_{ij} = (T_i - T_j)/R_{ij}.
\end{equation}
If nodes $i$ and $j$ are separated by a solid material, $R_{ij}$ is
given by \Cref{eq:R_from_k_eff}.
For a convective heat flow across a surface area $A$, heat flux is
given by \Cref{eq:htc_bc} and the corresponding thermal resistance is
\begin{align}
  \label{eq:R_conv}
  R_{\rm conv} &= 1/(h A).
\end{align}

The energy balance for node $i$ yields
\begin{align}
  \label{eq:network}
  \sum_{j} Q_{ij} \pp{T_i, T_j} &= 0,
\end{align}
where the sum is over all nodes $j$ connected to $i$.
\Cref{eq:network} represents a set of non-linear equations that can be
solved for the unknown node temperatures.

\subsubsection{Cold spot estimation}
\label{sec:cold_spot}

In addition to calculating heat ingress, a thermal model should be
able to describe cold spots in the structure. A cold spot may form,
e.g., where a metallic support structure intersects with an outer part
of the tank. Low cold-spot temperatures may result it ice formation,
or even condensation of air, and cold spots can be associated with
large temperature gradients and correspondingly high thermal stresses.
Furthermore, an accurate prediction of cold spots will improve the
accuracy of heat flow calculations for the thermal network model.

\begin{figure}[htbp]
  \includegraphics{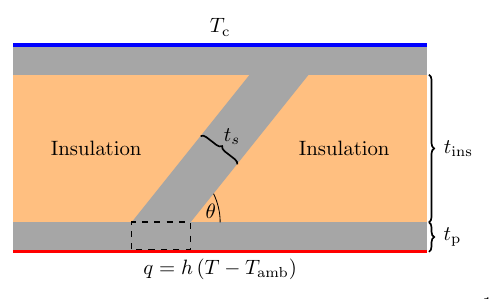}
  \caption{Schematic of the two-dimensional geometry considered in the
    cold spot estimation. Two parallel steel plates are connected by
    a supporting structure of thickness $t_s$, slanted at an angle
    $\theta$. The remainder is filled with insulation. A Dirichlet
    boundary condition is applied to the cold surface (blue line), and
    a convective boundary coefficient to the ambient is applied at the
    warm surface (red line). A cold spot will develop in
    the dashed region where the support meets the warm
    plate. \label{fig:validation-geometry}}
\end{figure}

A representative case that captures a source of cold spots is the
effectively two-dimensional geometry shown in
\Cref{fig:validation-geometry}: Two plates, one warm and one cold,
have an insulation material between them and are connected by a
support structure of thickness $t_{\rm s}$. The two plates are
separated by a distance $t_{\rm ins}$ and the warm plate has thickness
$t_{\rm p}$. To avoid considering boundary effects, we assume the
plates extend to infinity in the horizontal directions. The inside of
the cold plate is assumed to be at a constant temperature $\Tc$, while
the outside of the warm plate exchanges heat with the ambient
temperature so that the heat flux across it is given by
\eqref{eq:htc_bc}.

To a first approximation, the temperature in the warm plate is uniform
in the direction normal to the plate, and only a function $T\pp{\xi}$
of the distance $\xi$ along the plate, away from the cold spot.
Letting $T_\infty = \lim_{\xi\to\infty} T\pp{\xi}$ and $\Delta T
\pp{\xi} = T\pp{\xi} - T_\infty$, one can derive the approximation
\begin{align} \label{eq:T_cs}
  \Delta T\pp{\xi} &= \Delta T\pp{0} \exp \qty(-|\xi|/\Xi),  \\
  \Xi &= \sqrt{k_{\rm p}(T_\infty) t_{\rm p}/( h + k_\ins\pp{T_\infty}/t_\ins) },
\end{align}
where $k_{\rm p}$ and $k_\ins$ are the thermal conductivities of the
warm plate and the insulation, respectively and length scale $\Xi$
describes how big the cold spot will be. The heat flow through the
warm plate into the cold spot is given by
\begin{align}
  \label{eq:Q_cs}
  Q_{\rm cs} &= -\frac{\Delta T \pp{0}}{R_{\rm cs}},
\end{align}
where
\begin{align}
  \label{eq:R_cs}
  R_{\rm cs} &= \frac{\Xi}{2 L t_{\rm p} k_{\rm p} \pp{T_\infty}},
\end{align}
and $L$ is the length of the connection between the support and the
warm plate, i.e.\ the thickness of the two-dimensional geometry
depicted in \Cref{fig:validation-geometry}.

A cold-spot correction is added to the network model by introducing a
resistance given by \Cref{eq:R_cs}. We give a concrete example of its
application in \Cref{sec:case-study}, and we shall see that the
correction indeed gives more accurate heat flow estimates. In
addition, \Cref{eq:T_cs} can be used for estimating temperature
profiles near a cold spot. Both equations, however, require the
faraway temperature $T_\infty$, which can be calculated in different
ways. When a resistance $R_{\rm cs}$ is included in the network model,
$T_\infty$ is obtained as part of the solution. This results, however,
in a slight underestimation of $T_\infty$. To estimate the temperature
profile $T\pp{\xi}$ using \Cref{eq:T_cs}, we therefore first solve a
network model without support structure to get $T_\infty$.

Local corrections such as the one presented here to account for the
cold spot are a standard tool in thermal engineering that can be
derived for various geometries, e.g., to describe heat dissipation in
fins~\cite{incropera_ed6}. A detailed derivation and further
validation of accuracy can be found in the Supplementary Information
(SI).

\section{Numerical methods}
\label{sec:numerical}

In this section, we briefly describe the numerical methods used to
solve the thermal network model and the heat equation.

\subsection{Network model}
\label{sec:numerical-network}

The network model consists of a non-linear system of
equations~\eqref{eq:network} for the unknown node temperatures. These
were solved iteratively by successively solving linearized
approximations to \Cref{eq:network}, with constant resistances. In
each iteration, thermal resistances were updated using the current
node temperatures and then the linear system was solved using
\verb|DGESV| from the LAPACK library~\cite{laug}, called through the
NumPy~\cite{harris2020array}.  Iteration continued until the node
temperatures converged.

Once the network model is solved, temperature profiles $T\pp{x}$ along
individual resistances in the network may be obtained. This was here
accomplished by splitting the resistance under considerations into $N$
resistances in series. Then, starting at the cold side temperature,
\Cref{eq:Q_from_R} was solved for $\Delta T$ for each resistance in
succession, thus obtaining the temperature at the $N-1$ intermediate
nodes.

\subsection{Heat equation}

This section describes how we solve the steady-state heat
equation~\eqref{eq:heat-eqn} numerically using the finite element
method.  Let $\Omega$ be the domain on which we want to solve the
equation (i.e., the cryogenic tank, including insulation and support
structure),
and let $\Gamma_{\textsc{d}}$ and $\Gamma_{\textsc{r}}$
denote the parts of domain
boundary where we impose Dirichlet and Robin boundary conditions, respectively.
On the remainder of the boundary, we impose homogeneous Neumann conditions (i.e., zero heat flux).
For a given
simplicial mesh $\mathcal{T}_h = \pc{\tau}$ of $\Omega$ and prescribed temperature $T_{\textsc{d}}$ on $\Gamma_{\textsc{d}}$, we define the
trial function space $V_{h, T_{\textsc{d}}}={\pc{f \in
  C(\Omega)\ \big|\ f|_{\tau} \in \mathbb{P}^3(\tau)\ \forall \tau \in
  \mathcal{T}_h \text{ and } f|_{\Gamma_{\textsc{d}}}=T_{\textsc{d}}
  }}$ and its corresponding test function space $V_{h, 0}, $
where
$\mathbb{P}^3(\tau)$ is the space of cubic polynomials defined on the
mesh element $\tau$.
Then, the \textit{discrete weak formulation} of
\Cref{eq:heat-eqn}, supplemented with the Robin boundary condition of~\Cref{eq:htc_bc},
reads: 
Find 
$T_h \in V_{h,T_d}$ such that
\begin{equation}
  \int\limits_{\Omega}\! k\pp{T_h} \nabla T_h \cdot \nabla v_h \,
  \dd x +
   \!\! \int\limits_{\Gamma_{\textsc{r}}}\!\!\htc
  T_h v_h\, \dd s = 
   \!\! \int\limits_{\Gamma_{\textsc{r}}}\!\!\htc
  T_{\mathrm{amb}} v_h\, \dd s
  \label{eq:weak_form}
\end{equation}
for all test functions $v_h \in V_{h,0}$.\footnote{Since the
containment system is made from multiple materials, $k$ is
discontinuous across material interfaces.}  
We handle the non-linearity in \Cref{eq:weak_form} resulting from the
temperature dependence of $k$ through linearization and Newton
iterations. Mesh creation, linearization, assembly of a linear system
of equations representing \Cref{eq:weak_form}, and solution of the
assembled system of equations, are all handled within the open-source
finite element software suite
Netgen/NGSolve~\cite{schoberl1997netgen,schoberl2014c++},
which we access through its Python interface. Since the
tanks considered herein are rotationally symmetric, we express
\Cref{eq:weak_form} using cylindrical coordinates and use a
two-dimensional mesh (visualized in the Supplementary Information)
for increased computational efficiency.
The mesh consists of \SI{147 989}{} elements over which we define
third-order basis functions (cf.\ the definitions of $V_{h,T_d}$
and $V_{h,0}$ above).
Curved domain surfaces are approximated to the third order using
NGSolve's in-built functionality for curved element facets.

Given the computed temperature field $T_h$, we estimate the heat
ingress by
\begin{equation}
  Q_h = \int\limits_{\Gamma_{\mathrm{in}}} \mathcal{P}^1\pc{-k\pp{T_h}\nabla
  T_h \cdot \vec{n}} \dd s,
  \label{eq:FEM_heat}
\end{equation}
where $\mathcal{P}^1\pc{\cdot}$ denotes projection into the
space of continuous, elementwise linear functions,
$\Gamma_{\mathrm{in}}$ is the tank's inner surface, and $\vec{n}$ is
its surface normal vector pointing into the contained \lhto. 
The integral in
\Cref{eq:FEM_heat} was evaluated numerically using NGSolve.

\section{Implications of material properties on heat ingress}
\label{sec:results}

In this section, we explore selected aspects of the relationship
between material properties and heat ingress.
We evaluate the validity of the Concavity Hypothesis for a
selection of materials relevant to \lhto applications, and discuss
its implications for uncertainty in insulation performance predictions.
We also show that the material-specific dependence of apparent thermal conductivity
on temperature has important implications for overall thermal performance
and sensitivity to boundary conditions.

\begin{figure*}[htbp]
  \centering
  \includegraphics[width=1.3\columnwidth]{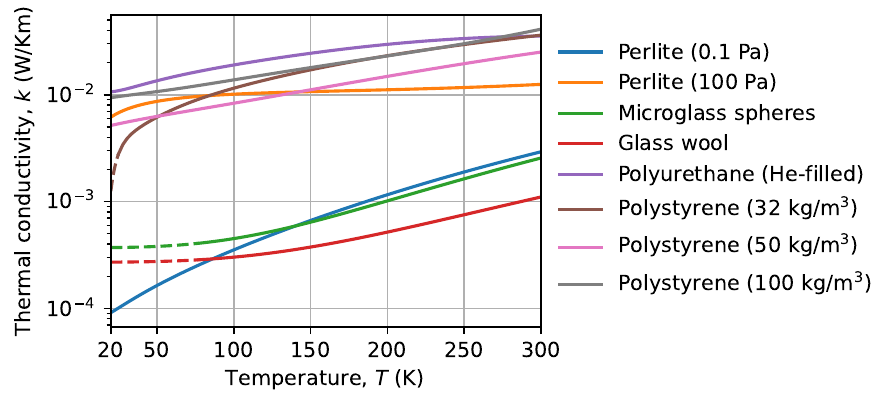}
  \caption{Thermal conductivities of insulation materials as a function of
    temperature, where the dashed portions represent extrapolations of the
    thermal conductivity correlations below their range of validity. Perlite
    correlations are from Ref.~\cite{ratnakar2023effective}, Microglass spheres
    and Glass wool correlations from Ref.~\cite{Hofmann2006}, and the remaining
    correlations from Ref.~\cite{NISTCryogenicsWebpage}.
    }
  \label{fig:k_curves_insulations}
\end{figure*}

\subsection{Validity of the Concavity Hypothesis}

The thermal conductivities $k\pp{T}$ of a number of insulation
materials are plotted in \Cref{fig:k_curves_insulations}. It is
readily observed that the thermal conductivities are all monotonically
increasing. This implies that the Concavity Hypothesis is valid for
these materials.
For the materials where $k\pp{T}$ is known reliably down to
\SI{20}{\kelvin}, we have verified this observation numerically by
calculating the thermal conductivity integral $K$ as well as the lower
and upper bounds from \Cref{sec:concavity}. We used \SI{77}{\kelvin}
as the intermediate temperature $\Tm$ when computing the bounds. The
results are shown in \Cref{tab:Kminmax-table}, where a selection of
structural materials, such as aluminum 5083 (Al5083) and stainless
steel 316 (SS316), are also included. For all the examples listed,
$K_{\rm min} \leq K \leq K^{\rm diff}_{\rm max} \leq K^{\rm int}_{\rm
  max}$, meaning that the Concavity Hypothesis is valid. It is also
evident that $K^{\rm diff}_{\rm max}$, in addition to being an upper
bound, is in fact an excellent approximation to $K$.

The Concavity Hypothesis is valid for many materials relevant to \lhto
applications, as demonstrated above. However, caveats are warranted
for three notable cases.

First, for some solid materials, thermal conductivity can have a
non-monotonic temperature dependence. A prominent example is pure
copper, which experiences a strong increase in thermal conductivity in
the \SI{20}{\kelvin}-range~\cite{simon1992copper}.  A similar effect
can also be observed in other pure metals, such as
aluminum~\cite{woodcraft2005aluminium}.
On the other hand, metallic alloys,
such as Al5083 and SS316 have numerous grain boundaries
that limit electronic heat transfer and are expected to have
monotonically increasing thermal conductivities~\cite{barron2017cryogenic}.
Hence, the hypothesis holds for these materials, as previously noted.

The second caveat pertains to porous insulation materials at
intermediate vacuum levels. The insulation materials in
\Cref{tab:Kminmax-table} are evaluated in either the high-pressure or
the low-pressure limit, which are the scenarios typically considered
in applications. As we have seen, the Concavity Hypothesis works well
in these scenarios. Moreover, the hypothesis' key assumption that
$k\pp{T}$ be monotonic is consistent with all published measurements
on insulation materials known to the authors. However, as exemplified
by the Ratnakar correlation~\cite{ratnakar2023effective}, thermal
conductivity can in principle become slightly non-monotonic in the
intermediate pressure range where gas conduction is strongly
pressure-dependent (the Knudsen regime). An example of this is shown
in the SI.

Finally, natural convection is a potential concern in non-evacuated
insulation materials. As temperature decreases from \ce{LN2} to \lhto,
the Rayleigh number increases and may exceed the critical Rayleigh
number for the establishment of convection cells~\cite{ratnakar2023short}.
Since the Rayleigh number is proportional to the characteristic length scale,
porous insulation materials with small void spaces are less susceptible
to this effect. Cracks and defects in the insulation, however, can promote
convection.

\begin{figure*}[htbp]
  \centering
  \includegraphics[width=\textwidth]{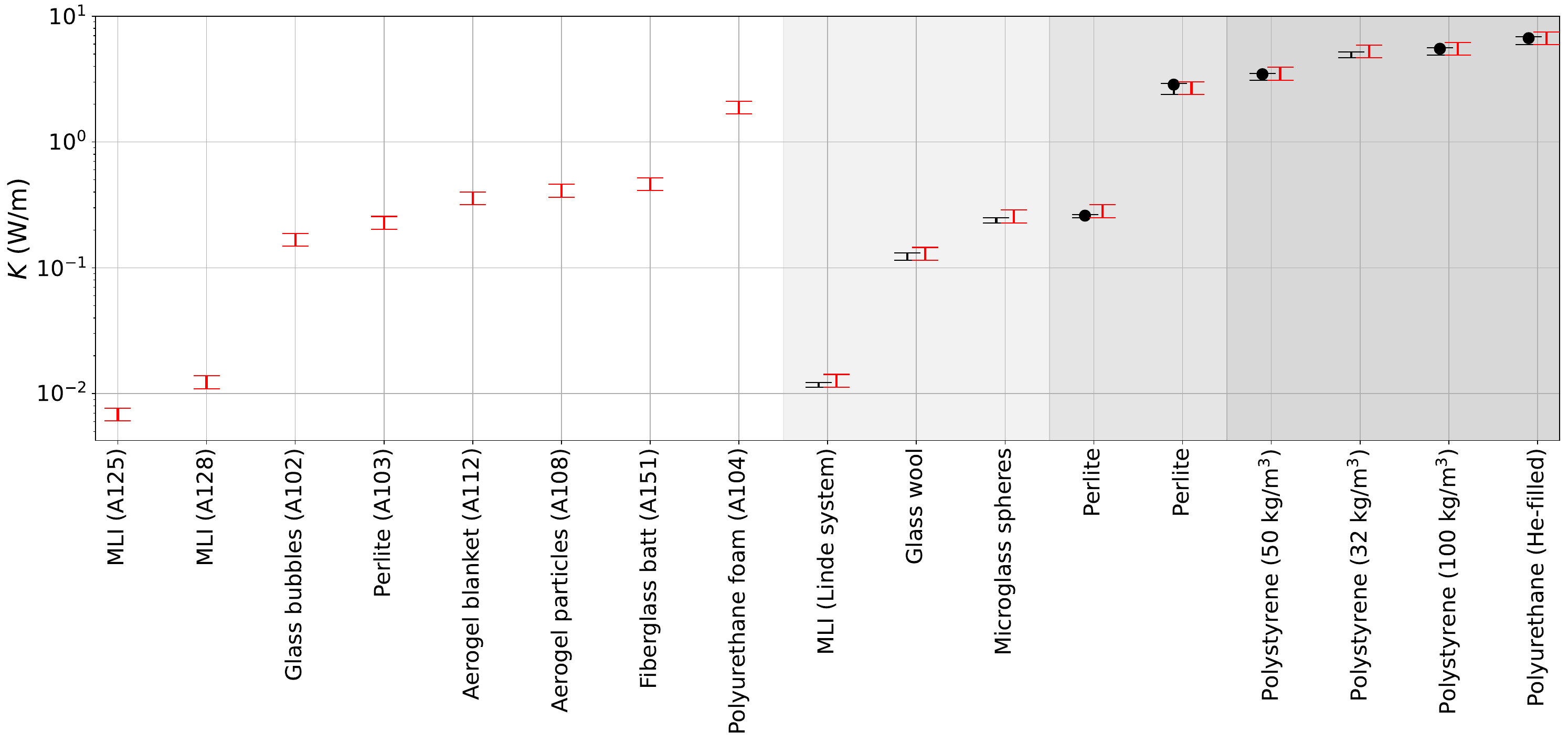}
  \caption{Thermal conductivity integral for insulation materials
    between \SI{20}{\kelvin} and \SI{293}{\kelvin}. The black dots
    correspond to the actual value of $K$ in the case it is known from
    measurements.
    The black interval corresponds to the
    bounds one can derive if the thermal conductivity is known \emph{at}
    \SI{77}{\kelvin}, whereas the red intervals correspond to the
    bounds one can derive from a single heat ingress measurement
    between \SI{77}{\kelvin} and \SI{293}{\kelvin}. The different
    background shades correspond to different references, which are, from
    left to right:~\cite{Fesmire2015},~\cite{Hofmann2006},
    ~\cite{ratnakar2023effective},~\cite{NISTCryogenicsWebpage}. The
    warm temperature \SI{293}{\kelvin} was chosen since this was the
    temperature used in the measurements by
    Fesmire~\cite{Fesmire2015}. The identifiers (A125 etc.) correspond
    to the indexing scheme by Fesmire~\cite{Fesmire2015}. }
  \label{fig:relative_thermal_performance_everything}
\end{figure*}

\subsection{Application of the Concavity Hypothesis}
\label{sec:concavity_application}

For the insulation materials considered in \Cref{tab:Kminmax-table},
both the thermal conductivity integrals $K \pp{ \SI{20}{\kelvin},
  \SI{293}{\kelvin}}$ and their corresponding bounds are visualized in
\Cref{fig:relative_thermal_performance_everything}.  The figure also
includes bounds, derived using the Concavity Hypothesis, for
insulation materials whose exact $K \pp{ \SI{20}{\kelvin},
  \SI{293}{\kelvin}}$ is not known.

\Cref{fig:relative_thermal_performance_everything} offers a robust
clarification of the relative performance of different insulation
materials. The performance differences can be gleaned even from the
red error bars alone, obtained by a single measurement between a CBT
of \SI{77}{\kelvin} and a WBT of \SI{293}{\kelvin}.  This indicates
that uncertainty due to the CBT of measurements is unlikely to
significantly affect the overall performance hierarchy.  Additionally,
as previously noted, the difference between $K$ (the black dots) and
$K^{\rm diff}_{\max}$ (the upper black error bar) is small. This
excellent agreement is quantified in \Cref{tab:Kminmax-table}, with relative errors below 5\%, and
indicates that $K^{\rm diff}_{\max}$ is a practically useful estimate
of $K$.

\begin{table*}[ht]
  \centering
    \caption{The thermal conductivity integral $K = K(\SI{20}{\kelvin},
    \SI{293}{\kelvin})$ for \lhto and ambient boundary temperatures, as well as
    the bounds calculated from information in the interval [77 K, 293
    K]. Here $K_{\min} = K(\SI{77}{\kelvin}, \SI{293}{\kelvin})$ and the
    quantities $K_{\rm max}^{\rm diff}$ and $K_{\rm max}^{\rm int}$ correspond
    to the theoretical upper bounds \Cref{eq:tight_bound} and \Cref{eq:loose_bound_K}, respectively.
    }
  \label{tab:Kminmax-table}
  \begin{tabular}[c]{lccccc}
    \toprule
    Material                  & $K_{\rm min}$ (\si{\W/m})   & $K$ (\si{\W/m})          & $K_{\rm max}^{\rm diff}$ (\si{\W/m})   & $K_{\rm max}^{\rm int}$ (\si{\W/m}) & Ref.\\
    \midrule
    Perlite (\SI{0.1}{\pascal})      & 0.251           & 0.260           & 0.265           & 0.317           & \cite{ratnakar2023effective} \\
    Perlite (\SI{100}{\pascal})      & 2.38            & 2.86            & 2.93            & 3.01            & \cite{ratnakar2023effective} \\
    Perlite (\SI{1}{\bar})       & 5.90            & 6.59            & 6.82            & 7.46            & \cite{ratnakar2023effective} \\
    Polyurethane (\ce{He}-filled)  & 5.93            & 6.69            & 6.87            & 7.49            & \cite{NISTCryogenicsWebpage} \\
    Polystyrene (100 \si{kg/m^3}) & 4.90            & 5.51            & 5.60            & 6.20            & \cite{NISTCryogenicsWebpage} \\
    Polystyrene (50 \si{kg/m^3})  & 3.10            & 3.46            & 3.52            & 3.92            & \cite{NISTCryogenicsWebpage} \\
    Aluminum 5083                 & 20000           & 22200           & 23100           & 25300           & \cite{NISTCryogenicsWebpage} \\
    Stainless steel 316            & 2600            & 2910            & 3050            & 3280            & \cite{NISTCryogenicsWebpage} \\
    Fiberglass (normal direction)  & 92.5            & 106             & 108             & 117             & \cite{NISTCryogenicsWebpage} \\
    Teflon                         & 56.4            & 67.9            & 69.6            & 71.3            & \cite{NISTCryogenicsWebpage} \\
    \bottomrule
  \end{tabular}
\end{table*}

Generally, when only a single measurement of heat flow through a
material is available, the Concavity Hypothesis can still provide
useful insights.  According to \Cref{eq:Qmaxint}, the global upper
bound only depends on the boundary temperatures. In particular, for
$\Tc=\SI{20}{\K}$, $\Tc=\SI{77}{\K}$, $\Tw=\SI{293}{\K}$, we obtain
that the heat ingress increases by no more than $\SI{26}{\percent}$
when lowering the CBT from \ce{LN2} to \lhto temperatures. The average
of the upper and lower bounds in \Cref{tab:Kminmax-table}, which
corresponds to a $\SI{13}{\percent}$ increase, can be used as a
rule-of-thumb for the increase in heat ingress when translating from
\ce{LN2} to \lhto applications when only a single heat ingress
measurement has been performed.

Another application of the Concavity Hypothesis is in the design of
experiments. These are often done using \ce{LN2} at its boiling point
as the CBT. It is, however, of interest to know the heat ingress for
lower temperatures as well. Although the \SI{13}{\percent} rule of
thumb can be used, $Q_{\max}^{\rm diff}$ is a better estimate that is
both accurate and also conservative (i.e.\ an overestimate). To
calculate $Q_{\max}^{\rm diff}$, it follows from \Cref{eq:Qmaxdiff}
that one must estimate $\partial Q(\Tm, T_{\amb})/\partial \Tm$. If an
experimental setup is used with \ce{LN2} at atmospheric pressure,
i.e. $\Tm=\SI{77}{\kelvin}$, one can obtain a slightly higher CBT by
increasing the pressure of the nitrogen in the tank. For \ce{LN2}, the
saturation temperature increases with roughly \SI{6}{\kelvin \per
  \bar}, and therefore another measurement with \SI{2}{\bar} internal
tank pressure may suffice to estimate the derivative.

\subsection{Sensitivity of heat ingress to boundary temperatures}
\label{sec:boundary-sensitivity}

One aspect that follows immediately from \Cref{eq:Q_from_K,eq:K-def}
is the sensitivity of $Q(\Tc,\Tw)$ to the boundary temperatures $\Tc$ and
$\Tw$. Differentiating \Cref{eq:Q_from_K} and dividing by $Q$, we get
that
the relative change in heat ingress is independent of the geometry,
and given by
\begin{align}
  \frac{1}{Q}\qty( \pderivative{Q}{\Tw} )_{\Tc} &= \frac{ k\pp{\Tw}}{K\pp{\Tc, \Tw}},
  \label{eq:dQdTh} \\
  \frac{1}{Q}\qty( \pderivative{Q}{\Tc} )_{\Tw} &= \frac{
    -k\pp{\Tc}}{K\pp{\Tc, \Tw}}.
  \label{eq:dQdTc}
\end{align}
Since the thermal conductivity of many materials is a strongly
increasing function of temperature, this implies that the WBT has a
much larger impact on the heat ingress than does the CBT.

\begin{figure}[htbp]
  \centering
  \begin{subfigure}{0.5\textwidth}
    \includegraphics[width=0.9\columnwidth]{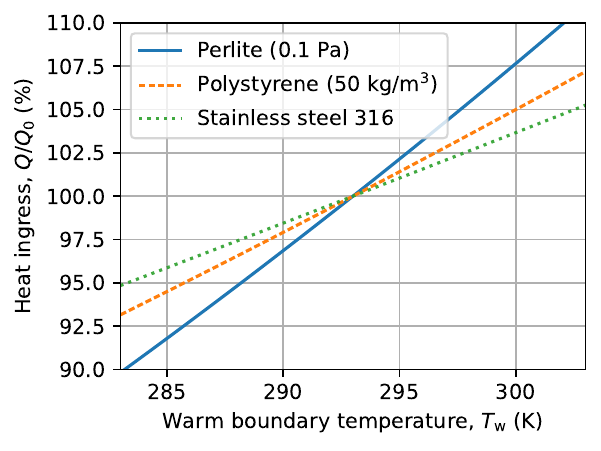}
    \caption{}
  \end{subfigure}
  \begin{subfigure}{0.5\textwidth}
    \includegraphics[width=0.9\columnwidth]{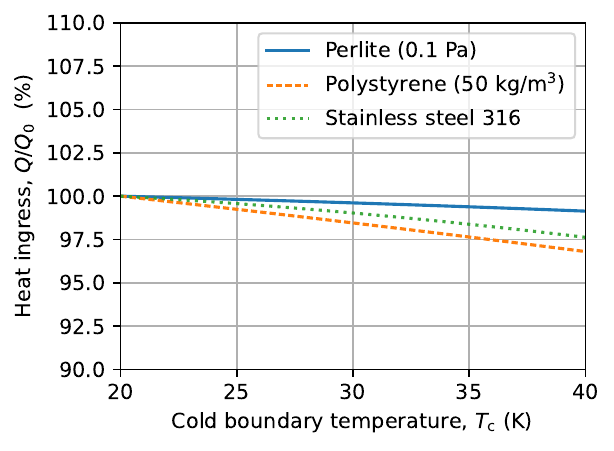}
    \caption{}
  \end{subfigure}
  \caption{Response in heat ingress $Q \pp{\Tw,\Tc}$ through three
    materials when (a) varying $\Tw$, keeping $\Tc=\SI{20}{\kelvin}$,
    and (b) varying $\Tc$, keeping $\Tw=\SI{293}{\kelvin}$. In both
    cases, the heat ingress $Q$ has been normalized by $Q_0 =
    Q\pp{\Tc=\SI{20}{\kelvin}, \Tw=\SI{293}{\kelvin}}$.}
  \label{fig:boundary-temp-sensitivity}
\end{figure}

\Cref{fig:boundary-temp-sensitivity} shows how heat ingress varies
with respect to changes in the WBT and CBT for rectangular slabs made
from two example insulation materials (evacuated perlite and
polystyrene) and a structural material (SS316).  As
expected from \Cref{eq:dQdTh,eq:dQdTc}, the heat ingress is much less
sensitive to the CBT than to the WBT. We find the same to be true also
for the other materials listed in \Cref{tab:Kminmax-table}. For perlite
the relative increase in heat ingress is on the order of 1\%
per Kelvin increase in WBT, whereas the sensitivity for the CBT is an order
of magnitude lower.

The observation above has at least three implications of great
practical importance. First, it means that ambient conditions should
be carefully considered when designing \lhto tanks and related
infrastructure. Secondly, it implies that comparatively low thermal
conductivity at higher temperatures is an attractive feature in
insulation materials, due to its large impact on $K$.  Thirdly,
thermal performance can only be estimated accurately if the warm
boundary surface temperature is suitably represented in the thermal
model used.  For thermal network models, this provides a clear
theoretical motivation for including the cold spot correction
presented in \Cref{sec:cold_spot}.  The practical benefit of this
correction is investigated in \Cref{sec:heat_cold_spot}.

\section{Case study of self-supported spherical tank}
\label{sec:case-study}

In this section we study the heat ingress, boil-off rate and
temperature distribution in a particular tank geometry. For this
purpose, we employ the network model, both with and without the
cold-spot correction, and the heat equation solved with the FEM. A
detailed discussion of the results and a comparison of the modeling
approaches is done in \Cref{sec:heat_cold_spot}. In
\Cref{sec:sensitivity} we consider the effect of changing the
dimensions of the support structure, the effect of changes in the
ambient temperature and the effect of choosing different insulation
materials.

The considered tank is intended for ship transport of \lhto and is a
\SI{40000}{\meter\cubed} spherical double-jacketed and vacuum
insulated tank supported with a cylindrical skirt. It is similar in
structure to the tank design developed by Moss Maritime, described in
the patent application in Skogan et al.~\cite{Skogan2022}. The
specific dimensions of tank components chosen here are considered to
be structurally viable, but not necessarily optimal, for an \lhto
tank.

\begin{figure}[htbp]
  \centering
  \includegraphics[trim=30 0 40 0,width=0.96\columnwidth]{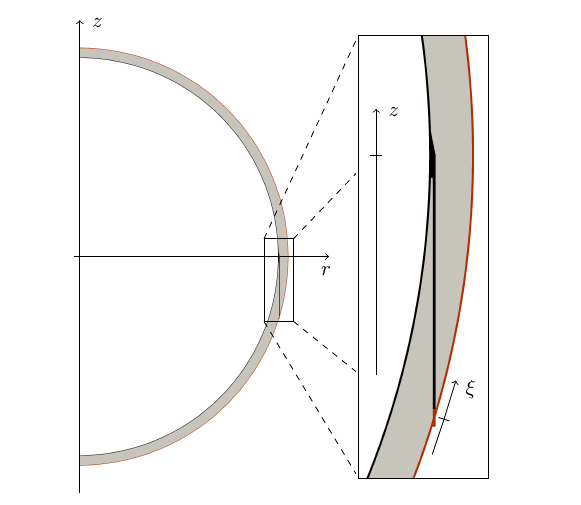}
  \caption{Illustration of the spherical \lhto tank. The inner tank,
    the equatorial ring and support skirt are made of stainless steel
    316 (black), the outer tank and mounting ring of carbon steel
    (red) and the annulus is filled with insulation material
    (gray). The inset shows a close-up of the support skirt. The
    $z$-coordinate indicates vertical position, with $z=0$ at the
    equatorial line. The coordinate $\xi$ denotes the distance along the
    outer tank wall from the intersection between the skirt and the
    outer tank.}
  \label{fig:moss_case_fem}
\end{figure}

The tank consists of four basic parts: (1)~A spherical inner tank made
from stainless steel 316 (SS316) with internal radius of
\SI{21.2}{\meter} and a thickness of \SI{5}{\centi\meter}, (2)~a
spherical outer tank made from carbon steel with internal radius
\SI{22.25}{\meter} and a thickness of \SI{5}{\centi\meter}, (3)~a
cylindrical support skirt with a thickness of $t_\supp =
\SI{6.5}{\centi\meter}$ made from SS316 that connects the equatorial
part of the inner tank to the outer tank and (4)~an annular space in
between the inner and outer tank that contains insulation material. In
addition, the skirt is connected to the inner tank by an equatorial
ring and to the outer tank by a cylindrical mounting ring. The latter
mounting ring extends slightly outside the outer tank to form an
interface towards the ship's inner hull. The geometry is rotationally
symmetric and is illustrated in \Cref{fig:moss_case_fem}. The mesh
used when solving the heat equation with the FEM is illustrated in the
SI.

Stainless steel is considered a relevant example material for the
inner tank and the support skirt. It is a class of alloys viewed as
compatible with hydrogen~\cite{SanMarchi2012}, and it has been used
successfully by NASA in their \lhto storage tanks at Kennedy Space
Center~\cite{Krenn2012}. Furthermore, it is a deliberate choice to
consider a support skirt made of the same material as the inner
tank. This eliminates the need to have joints between different
materials in the coldest regions of the tank, which may be problematic
due to different thermal expansion
properties~\cite[p.\ 129]{Bostock2019}.

In calculations, the thermal conductivity of SS316 is modeled by
\cref{eq:k_NIST} with parameters from~\cite{NISTCryogenicsWebpage},
while the thermal conductivity of carbon steel is set to
\SI{51.9}{\watt\per\meter\per\kelvin}.  Furthermore, the temperature
of the inner wall of the inner tank is kept constant at $\Tc =
\SI{20}{\kelvin}$. At the outer wall of the outer tank, a convective
boundary condition is used, with heat transfer coefficient $\htc =
\SI{2.5}{\watt\per\meter\squared\per\kelvin}$ and $T_\amb =
\SI{293}{\kelvin}$. The interface to the ship hull is assumed
adiabatic.

From the heat ingress $Q$ into a cryogenic storage tank one may
estimate the boil-off rate (BOR) by dividing by the latent enthalpy of
vaporization. This implicitly assumes no superheating of the gas that
is let out of the tank to keep pressure constant at 1 atm. The BOR is commonly
measured as the mass of liquid evaporated per time relative to the
mass of liquid in a full tank,
\begin{equation}
  \label{eq:BOR}
  \mathrm{BOR} = \frac{Q}{\rho V \Delta H},
\end{equation}
where $\rho$ is the liquid density, $V$ is the tank volume, and
$\Delta H$ is the specific latent heat. The typical unit is percent
per day (\si{\percent\per\day}).

\subsection{Heat ingress and cold spot calculation}
\label{sec:heat_cold_spot}

In this section, we calculate the heat ingress and temperature
distribution for the tank geometry described in the previous section,
with perlite insulation at a pressure of \SI{e-2}{\pascal}. We employ
the network model, both with and without the cold-spot correction, and
the heat equation solved with the FEM. The same tank geometry is
represented in all three models, the same boundary conditions are
applied and the same material thermal conductivity models are
used. The thermal conductivity of perlite is modeled by
\Cref{eq:k_ratnakar} with parameters
from~\cite{ratnakar2023effective}.

While the network model represents a simplified version of the
geometry, the heat equation resolves additional details. In
particular, the equatorial ring that connects the inner tank with the
skirt and the mounting ring are not included in the network model, and
the height $L_\supp = \SI{5.672}{\meter}$ of the skirt in the network
model is the corresponds to the distance between the equatorial ring
and the mounting ring.

\begin{figure}[htbp]
  \centering
  \begin{subfigure}{0.5\textwidth}
    \includegraphics[width=0.9\columnwidth]{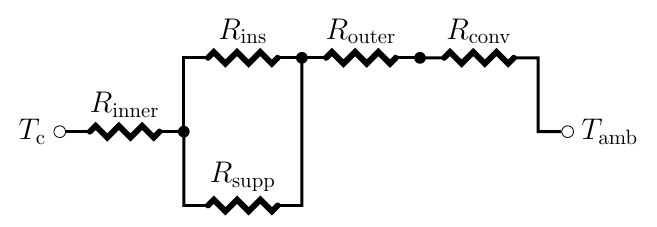}
    \caption{}
  \end{subfigure}
  \begin{subfigure}{0.5\textwidth}
    \includegraphics[width=0.9\columnwidth]{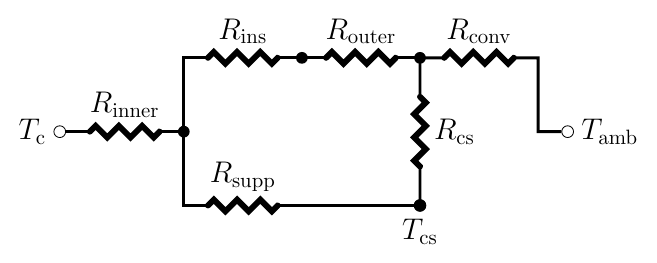}
    \caption{}
  \end{subfigure}
  \caption{Representations of the spherical \lhto tank in (a) the
    uncorrected network model and (b) the corrected network model. In
    both networks, each node represents a temperature. The white nodes
    have specified temperatures while the model needs to be solved to
    get the temperatures in the black nodes. In (b), the thermal
    resistance $R_{\rm cs}$ corresponds to heat flow into the cold
    spot.}
  \label{fig:circuit}
\end{figure}

The networks used are illustrated in \Cref{fig:circuit}. The
resistances $R_{\rm inner}$, $R_{\rm ins}$, $R_{\rm supp}$ and $R_{\rm
  outer}$ were calculated using \Cref{eq:R_from_k_eff} and represent,
respectively, the inner tank, the insulation, the support skirt and
the outer tank. The resistance $R_{\rm conv}$ represents convective
resistance to the ambient (see \Cref{eq:R_conv}) and $R_{\rm cs}$
models heat flow into the cold spot (see \Cref{eq:R_cs}).

The resulting heat ingress and cold-spot temperature from the three
different models are given in \Cref{tab:base_case}. It can be seen
that the uncorrected network model over-predicts the heat ingress by
\SI{2.8}{\percent}, w.r.t.\ the results from the heat equation. By
accounting for the cold-spot, however, the over-prediction is reduced
by one order of magnitude, to \SI{0.3}{\percent}.

The BOR is estimated to be approximately
\SI{0.04}{\percent\per\day}. This is in the lower range of what is
assumed in many high-level value chain studies in the literature,
which ranges from \SI{0.03}{\percent\per\day} to
\SI{1.35}{\percent\per\day}~\cite{berstad2022liquid}, It is, however,
in line with what can be expected when the benefit of scale,
i.e.\ small surface-to-volume ratio, is combined with the assumption
that the same vacuum conditions are attainable for the large tank under
consideration here as in smaller tanks currently in
operation~\cite{berstad2022liquid,Ratnakar2021}.

The uncorrected network model does not incorporate the effect of the
cold spot and therefore, as can be seen in \Cref{tab:base_case}, does
not predict any reduction in temperature at the cold spot. The
corrected network model, on the other hand, accurately reproduces the
cold-spot temperature from the heat equation. From
\Cref{tab:base_case}, the deviation in cold-spot temperature between
the heat equation and the corrected network model is only
\SI{0.6}{\kelvin}.

\begin{table*}[h]
  \centering
  \caption{Results for total heat ingress $Q$, heat ingress through
    support $Q_\supp$, cold spot temperature $T_{\coldspot}$ and BOR
    for the network model (corrected and uncorrected) and heat
    equation (FEM).}
  \label{tab:base_case}
  \begin{tabular}[c]{l r r r r}
    \toprule
    & $Q$ (\si{\kilo\watt})
    & $Q_\supp$ (\si{\kilo\watt})
    & $T_\coldspot$ (\si{\kelvin})
    & $\bor$ (\si{\percent\per\day})
    \\
    \midrule
    Network (uncorr.)
    & \SI{5.953}{}
    & 4.444
    & 292.6
    & 0.0405 \\
    Network (corr.)
    & \SI{5.809}{}
    & 4.298
    & 286.6
    & 0.0395 \\
    Heat eqn.
    & \SI{5.789}{}
    & -
    & 287.2
    & 0.0394 \\
    \bottomrule
  \end{tabular}
\end{table*}

\begin{figure}[htbp]
  \centering
  \begin{subfigure}{0.5\textwidth}
    \includegraphics[width=0.9\textwidth]{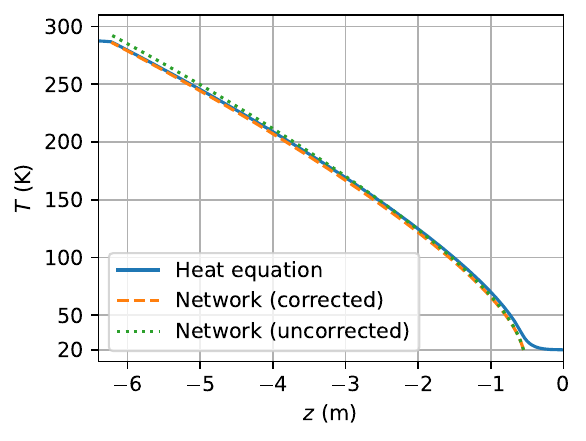}
    \caption{}
    \label{fig:base_case_fem_skirt}
  \end{subfigure}
  \begin{subfigure}{0.5\textwidth}
    \includegraphics[width=0.9\columnwidth]{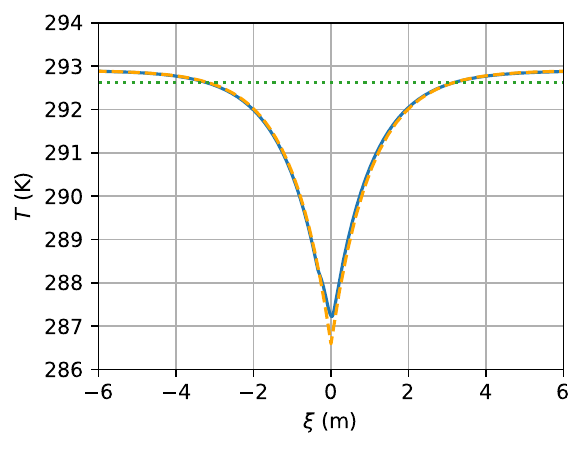}
    \caption{}
    \label{fig:base_case_fem_mantle}
  \end{subfigure}
  \caption{Temperature distribution in (a) the support skirt as a
    function of vertical position $z$ and (b) the outer tank as a
    function of distance $\xi$ from the cold spot. In both plots,
    results from the heat equation (FEM) are shown in solid blue,
    results from the corrected network model in dashed orange and
    results from the uncorrected network model in dotted green. }
\end{figure}

Three temperature profiles in the support skirt, along the vertical
direction, are shown in \Cref{fig:base_case_fem_skirt}. These profiles
were computed with, respectively, the heat equation, the corrected
network model and the uncorrected network model.
All three curves have a steeper gradient at cold temperatures due to
the lower thermal conductivity of SS316 in this region. Furthermore,
all three models roughly agree on the temperature variation along the
skirt. However, the two network models do not accurately reproduce the
heat equation profile near the innermost edge of the skirt where the
skirt in connected to the support ring. This is a result of the
simplified treatment of the structural geometry by the network model
in this region. Near the outermost end of the skirt, the corrected
model agrees very well with the heat equation results, while the
uncorrected model does not.

Three temperature profiles in the outer tank, as a function of
distance from the point where the outer tank and the skirt intersects,
are shown in \Cref{fig:base_case_fem_mantle}. These profiles were
again computed with the heat equation, the corrected network model and
the uncorrected network model, respectively. The corrected network
model captures not only the cold sport temperature, but also the
variation in outer tank temperature and the temperature at the cold
spot very well.

From the results presented in this section, we conclude that the
corrected network model accurately reproduces the results from full
solutions of the heat equation for the heat ingress, the cold spot
temperature and the variation in temperature along the outer tank
wall.

\subsection{Sensitivity analysis}
\label{sec:sensitivity}

In this section we use the three methods for heat ingress calculation
to study the effect of three important parameters: (1) the skirt
thickness, (2) the ambient air temperature and (3) choice of
insulation material. We thus assess the ability of the two variants of
the network model to predict how the heat ingress depends on these
parameters. In the case of insulation material, we also illustrate the
uncertainty in heat ingress resulting from using insulation thermal
conductivity measurements and models that are not valid for CBTs down
to \SI{20}{\kelvin}.

\paragraph{Effect of skirt dimensions}

For the case discussed in the preceding section, approximately
\SI{70}{\percent} of the heat ingress was transmitted through the
support skirt. It is therefore of interest to consider how the heat
ingress varies with skirt dimensions.

\begin{figure}[htbp]
  \centering
  \includegraphics[width=\columnwidth]{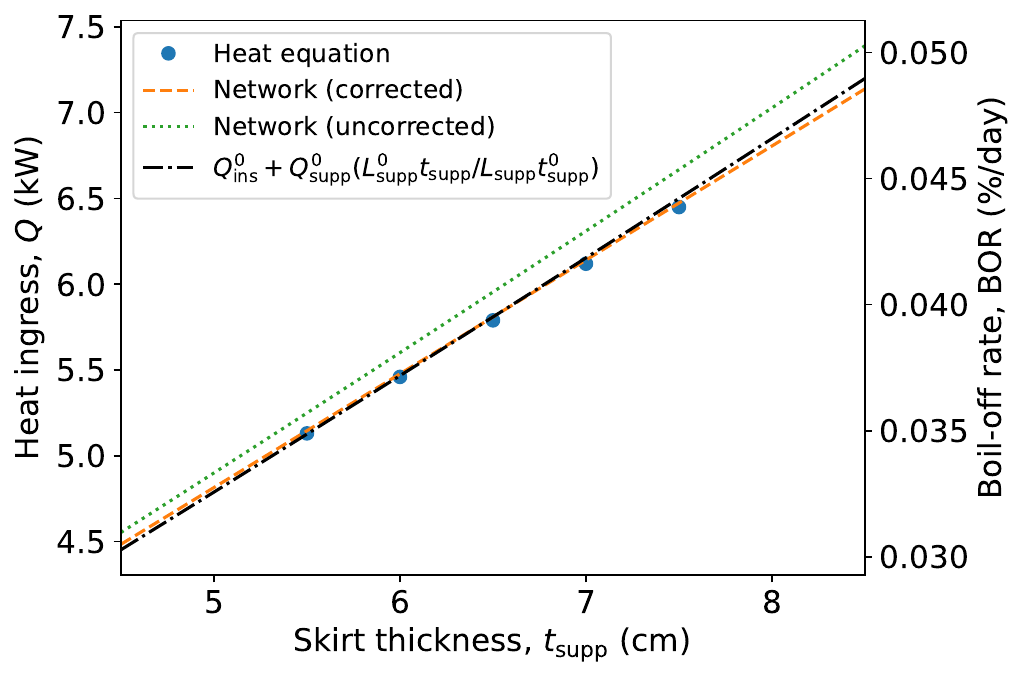}
  \caption{Variation in heat ingress and BOR with thickness $t_\supp$
    of the support skirt in the \lhto tank case. The increase in skirt
    thickness is also associated with a slight reduction of skirt
    height $L_\supp$. Results are obtained with the heat equation
    (blue dots), the network model with cold-spot correction (dashed
    orange) and the network model without cold-spot correction
    (dotted green). In addition, the total heat ingress and BOR
    obtained when scaling the heat flow $Q^0_\supp$ through a support
    of thickness $t_\supp^0 = \SI{6.5}{\centi\meter}$ and length
    $L^0_\supp$ with the factor $L^0_\supp t_\supp / L_\supp t^0_\supp
    $ is shown (dash-dotted black). The insulation heat flow
    $Q^0_\ins$ is calculated with support dimensions $t_\supp^0$ and
    $L_\supp^0$.}
  \label{fig:skirt_thickness}
\end{figure}

The variation of heat ingress and BOR with skirt thickness is shown in
\Cref{fig:skirt_thickness} for calculations performed with the heat
equation and the corrected and uncorrected network models. It is seen
that the total heat ingress increases approximately
\SI{0.7}{\kilo\watt} per \si{\centi\meter} of skirt thickness. This
increase is accurately captured by the corrected network model, while
the uncorrected network model over-predicts the heat ingress.

In varying the skirt thickness $t_\supp$, the inner radius of the
skirt is kept fixed and the outer radius is changed. This leads to a
slight reduction in the skirt height $L_\supp$ as the thickness is
increased. In \Cref{fig:skirt_thickness} the result of scaling the
heat ingress in the support with the skirt thickness $t_\supp$,
and the inverse of the skirt height $L_\supp$, is shown as a
dash-dotted black line. This line agrees reasonably well with both the
corrected network model and the heat equation
results. This agreement shows that most of the change in heat ingress
can be attributed directly to the resulting change in shape factor $S$
in the skirt, while the resulting changes in boundary temperatures and
the thermal conductivity integral $K$ are less important.

\paragraph{Effect of ambient temperatures}

As we showed in \Cref{sec:boundary-sensitivity}, the heat ingress
through both insulation and structural materials in \lhto tanks may be
sensitive to ambient temperatures. For a ship traversing the globe,
the ambient temperatures will vary along its route and it is therefore
important to consider the effect of ambient temperatures on heat
ingress and BOR. 

\begin{figure}[htbp]
  \centering
  \includegraphics[width=\columnwidth]{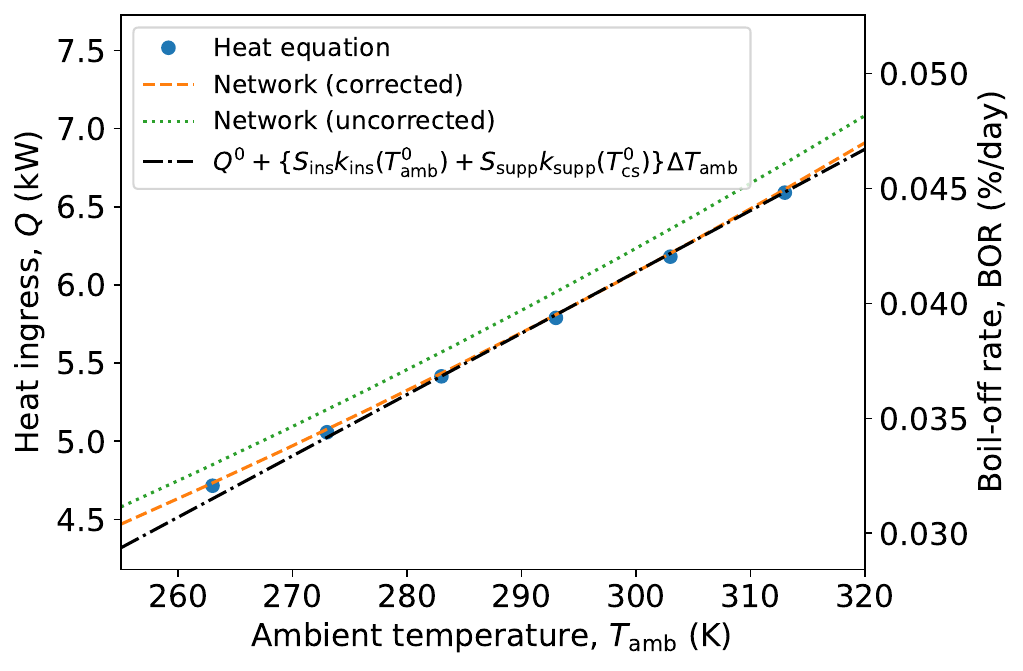}
  \caption{Variation of heat ingress and BOR with ambient
    temperature $T_\amb$ in the \lhto tank case. Results are obtained
    with the heat equation (blue dots), the network model with
    cold-spot correction (dashed orange) and the network model without
    cold-sport correction (dotted green). In addition, the heat
    ingress obtained when applying \Cref{eq:dQdTh} to the insulation
    and the support is shown (dash-dotted black). Herein, $Q^0$ is the
    total heat ingress and $T_{\rm cs}^0$ the cold sport temperature
    at $T_\amb^0 = \SI{293}{\kelvin}$. Furthermore, $S_{\ins}$ is the
    shape factor of the insulation, $S_\supp$ is shape factor of the
    support, $k_\ins$ is the thermal conductivity of the insulation,
    $k_\supp$ is the thermal conductivity of the support and $\Delta
    T_\amb = T_\amb - T^0_\amb$.}
  \label{fig:T_amb}
\end{figure}

The variation of heat ingress and boil-off rates with ambient
temperature are shown in \Cref{fig:T_amb}. The heat ingress and BOR
changes considerably, by approximately \SI{30}{\percent}, when the
ambient temperature increases from \SI{0}{\celsius} to
\SI{40}{\celsius}. As was the case when varying skirt dimensions, the
corrected network model accurately reproduces the results from the
heat equation, while the uncorrected network model over-predicts heat
ingress and boil-off rates.

The changes in ambient temperature do not affect any of the shape
factors in the network model, only the thermal conductivity integrals
of the different resistances. According to \Cref{eq:dQdTh}, this
variation can be approximated by the thermal conductivity evaluated at
the changing boundary temperature multiplied with the perturbation in
that temperature. Using this approximation, for both the insulation
and the support yields the black dash-dotted line in
\Cref{fig:T_amb}. This accurately reproduces the smaller perturbations
in ambient temperature. It is less accurate for larger perturbations
with a visible difference for changes larger than approximately
\SI{10}{\kelvin}.

The changes in ambient conditions in this section is limited to the
effect of convective heat transfer between the outside of the tank and
the ambient temperatures. In addition, there would be varying influx
of solar radiation due to changing sky conditions. We do not consider
such effects here, but only mention that they are expected to further
exacerbate the fluctuations in heat ingress and BOR.

\paragraph{Effect of different insulation materials}

An important factor in the design of a cryogenic storage tank is, of
course, choice of insulation material. In this section, we use data
and models from different literature
sources~\cite{ratnakar2023effective,Fesmire2015,Hofmann2006,NISTCryogenicsWebpage}
for the thermal properties of various insulation material and
calculate the effect on heat ingress and BOR in the \lhto tank
case. Calculations were performed with the corrected network
model. For the cases involving MLI, we have assumed that the thickness
of insulation is the same that is given in the original sources. The
insulation materials
from~\cite{ratnakar2023effective,Fesmire2015,Hofmann2006} are here
considered for pressures below $\SI{1.3e-2}{\pascal}$.

\begin{figure*}[htbp]
  \centering
  \includegraphics[width=0.95\textwidth]{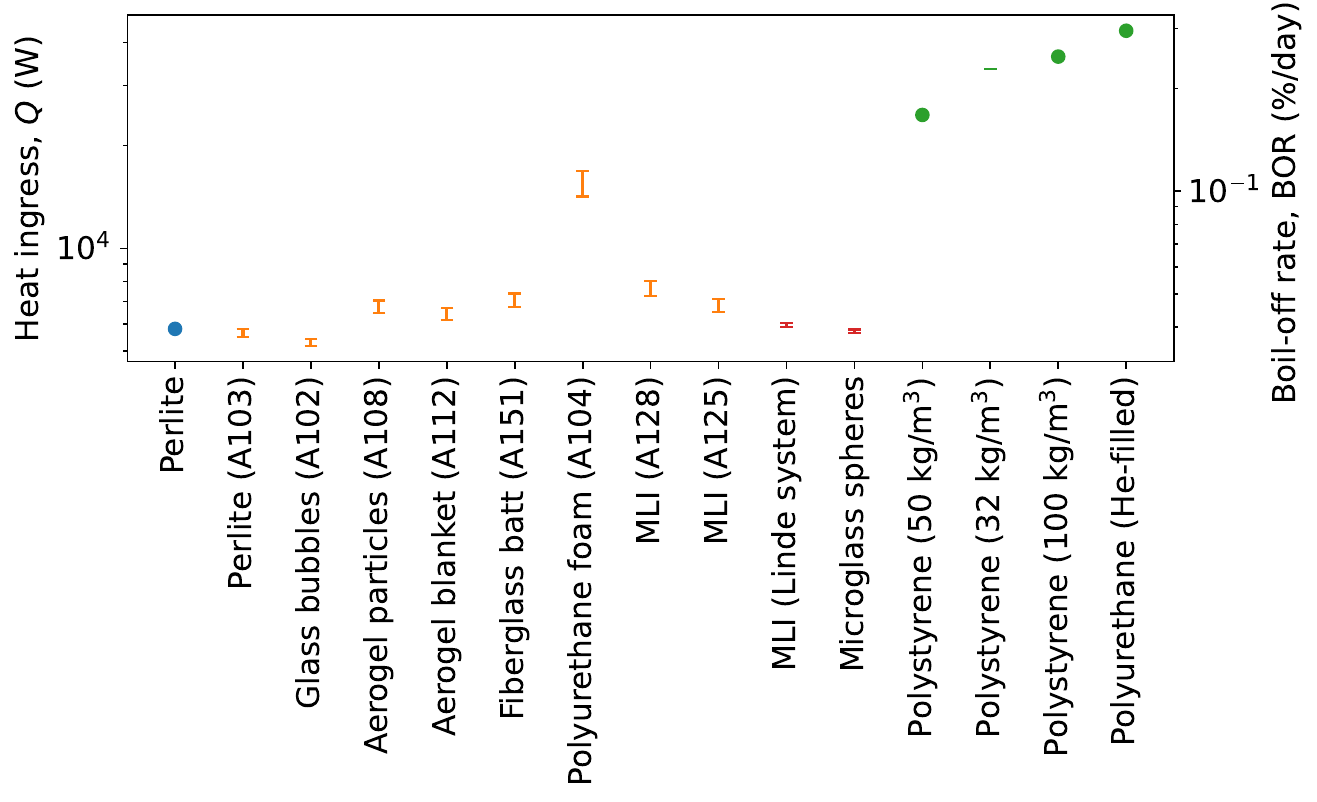}
  \caption{Heat ingress and BOR resulting from different choices of
    insulation materials in the \lhto tank case. Insulation thermal
    conductivity data/models are from~\cite{ratnakar2023effective}
    (blue),~\cite{Fesmire2015} (orange),~\cite{Hofmann2006} (red) and~\cite{NISTCryogenicsWebpage} (green). The error bars represent the
    uncertainty in the total heat ingress resulting from
    using insulation thermal conductivity measurements that have been
    performed with \SI{77}{\kelvin} rather than \SI{20}{\kelvin} as
    CBT. The WBT is 293 K. The identifiers (A103 etc.) correspond
    to the indexing scheme by Fesmire~\cite{Fesmire2015}.}
  \label{fig:tank_ins}
\end{figure*}

The variation of heat ingress and boil-off rates from different
choices of insulation material is shown in \Cref{fig:tank_ins}. The
various insulation materials correspond to a large range of boil-off
rates, from the lowest \SI{0.036}{\percent\per\day} (glass bubbles) to
\SI{0.3}{\percent\per\day} (Helium-filled polyurethane foam).

NASA has made extensive studies of cryogenic insulation performance,
much of which is summarized in~\cite{Fesmire2015}. The measurements
presented there were all performed with \SI{77}{\kelvin} as the
CBT. To use these measurements in the present case, with a cold CBT of
\SI{20}{\kelvin}, leads to some uncertainty in the heat ingress and
BOR. This uncertainty was evaluated using \Cref{eq:loose_bound}, and
is illustrated as orange error bars in \Cref{fig:tank_ins}. In
contrast, Hofmann~\cite{Hofmann2006} present thermal conductivity
models that are stated to be valid down to CBTs of
\SI{77}{\kelvin}. Their corresponding uncertainty when reducing the
CBT to \SI{20}{\kelvin} were calculated using \Cref{eq:tight_bound}
and are illustrated with red error bars in \Cref{fig:tank_ins}. From
this figure, it is evident that knowing $k$ at \SI{77}{\kelvin}, and
thus being able to use $K^{\rm diff}_{\max}$ (see \Cref{eq:tight_bound})
as upper bound rather than $K^{\rm int}_{\max}$ (see
\Cref{eq:loose_bound}), significantly reduces the uncertainty. The
models from~\cite{NISTCryogenicsWebpage}
and~\cite{ratnakar2023effective} are valid down to \SI{20}{\kelvin},
or close to it, and thus have no or negligible uncertainty.

NASA has published several papers comparing the thermal performance of
perlite powder and glass bubbles, see
e.g.~\cite{Fesmire2015,Sass2010}, and have found that glass bubbles
perform significantly better. Sass et al.\ found a reduction in \lhto
BOR of \SI{34}{\percent} in \SI{1}{\meter\cubed} demonstration tanks
when perlite was replaced with glass bubbles~\cite{Sass2008} and found
a reduction of \SI{44}{\percent} in a field-scale
tank~\cite{Sass2010}. In the present case, the reduction in heat
ingress through the insulation is smaller, approximately
\SI{27}{\percent}. Furthermore, this case also has a significant heat
ingress through the support skirt which is largely unaffected by the
choice of insulation material. The result is that the reduction in
total heat ingress and BOR (see \Cref{fig:tank_ins}) is only
\SI{7}{\percent}.

For most insulation materials in \Cref{fig:tank_ins} there is little
or no overlap of the error bars. Thus, the uncertainty introduced by
using thermal performance measurements with \SI{77}{\kelvin} as the
CBT in the present case is of limited importance for the qualitative
comparison of material performance in the tank considered. However,
the error bars may still represent a significant uncertainty in the
values of the various heat ingress and boil-off rates. For the
materials from~\cite{Fesmire2015} they are in the range of
\SI{4}{}--\SI{16}{\percent}, while for those from~\cite{Hofmann2006}
it is only \SI{2}{\percent}. For those from~\cite{NISTCryogenicsWebpage},
uncertainties are negligible.

One way to reduce the uncertainty is to obtain accurate thermal
properties also in the temperature region between \SI{77}{\kelvin} and
\SI{20}{\kelvin} for insulation materials where they are not
available. However, for the results from~\cite{Fesmire2015},
uncertainty can also be reduced as described in section
\Cref{sec:concavity_application}, without the need to do experiments
with a CBT of \SI{20}{\kelvin}.

The rightmost point in \Cref{fig:tank_ins} corresponds to
Helium-filled polyurethane foam. This is an unevacuated insulation
material with a fill gas that does not condense or freeze at \lhto
temperatures. Evacuated insulation materials exhibit excellent thermal
performance, but requires a structure that can hold the vacuum and
withstand buckling induced by the ambient pressure. Using
non-evacuated insulation materials has many advantages, as
pointed out in~\cite{ratnakar2023short}. One advantage is that the outer tank would then face a smaller
pressure difference between the insulation and the ambient, and
thus be less prone to buckling. The outer tank may then be made from a
thinner layer of material. This can particularly be important for
large tanks since they can become heavy if they rely on an evacuated
double-wall structure~\cite{Ratnakar2021}. Other advantages could
include improved safety, reduced cost, and no time needed for establishing a vacuum
by pumping. As we see from \Cref{fig:tank_ins}, however, this comes
at a cost of higher heat ingress and BOR. Compared to the base case of
evacuated perlite, the BOR is more than seven times higher. Still, the
rate of \SI{0.3}{\percent\per\day} may be within the range that could
be utilized for ship propulsion~\cite{ratnakar2023short}.

\section{Summary and conclusions}

Measurements of heat transfer through cryogenic insulation materials
are often made with liquid nitrogen (\ce{LN2}) at \SI{77}{\kelvin} as
the cold boundary temperature (CBT). This work addressed the
uncertainty in thermal performance if these materials are applied to
liquid hydrogen (\lhto) storage, where the CBT is often around
\SI{20}{\kelvin}. Specifically, we introduced the Concavity
Hypothesis, which states that the heat ingress is a concave,
decreasing function of the CBT. The Concavity Hypothesis relies on the
observation that thermal conductivity tends to be monotonically
increasing with temperature. This assumption was tested and found to
be valid for a number of available thermal conductivity correlations
that are valid down to \SI{20}{\kelvin}, and we have argued that it is
expected to be generally valid for evacuated insulation materials.

Assuming a warm boundary temperature of $\sim\! \SI{293}{\kelvin}$,
changing the CBT from \SI{77}{\kelvin} to \SI{20}{\kelvin} resulted in
an increase in the heat ingress from the ambient below
20\% in materials where reliable thermal conductivity
models are available (cf. \Cref{tab:Kminmax-table}). Using the Concavity Hypothesis, we derived an
upper bound, $Q^{\rm int}_{\max}$, of \SI{26}{\percent} increase, and
suggested using half of this (\SI{13}{\percent} increase) as a
rule-of-thumb. A tighter upper bound, $Q^{\rm diff}_{\max}$, is
available if one knows the material's thermal conductivity at
$\SI{77}{\kelvin}$. We found that this bound may serve as an accurate
\textit{estimate} of the heat ingress, with relative errors below
5\%. This clearly motivates the measurement of thermal
conductivity at $\SI{77}{\kelvin}$, and we outlined a possible
experimental procedure to do so.

We found that the heat transfer through a material is often much more
sensitive to perturbations in the temperature on the warm side than on
the cold side. This means that, when selecting \lhto insulation
materials, low thermal conductivity at high temperatures is
decisive. Using perlite as an example, we found that for small variations
in the ambient temperature, the heat ingress increased with as much as
\SI{1}{\percent} per Kelvin. This insight motivated studying the
effect on heat transfer through cold spots on the outside of tanks
introduced e.g.\ by structural support elements.

We next employed a thermal network model to analyze a cryogenic
storage tank designed for the large-scale maritime transport of
\lhto. The network model was enhanced with a correction to account for
the impact of external cold spots on the total heat ingress and to
assess the temperature distribution near such cold spots. The model's
predictions were validated against finite element method (FEM)
solutions of the steady-state heat equation, showing excellent
agreement with errors in total heat ingress and cold spot temperatures
within \SI{0.3}{\percent} and \SI{0.6}{\kelvin}, respectively.

Both the network model and the FEM solutions of the heat equation were
used to quantify the effects of support thickness and ambient
temperatures on heat ingress and boil-off. The network model, with the
cold spot correction, was shown to predict these variations well. A
notable finding was a potential increase in boil-off rates up to
\SI{30}{\percent} under plausible ambient temperature changes.

Finally, we used the network model to assess the effect of using
different insulation materials for the considered tank. While
evacuated insulation materials achieve the lowest boil-off,
non-evacuated insulation systems nevertheless appear promising if the
boil-off gas could be utilized for e.g.\ ship propulsion.

\section*{Acknowledgments}
This publication is based on results from the research project
\textit{\lhto Pioneer -- Ultra-insulated seaborne containment system
  for global \lhto ship transport}, performed under the ENERGIX
program of the Research Council of Norway. The authors acknowledge
the following parties for financial support: Gassco, Equinor, Air
Liquide, HD Korea Shipbuilding \& Offshore Engineering, Moss Maritime
and the Research Council of Norway (320233).

\section*{CRediT authorship contribution statement}

\textbf{Ailo Aasen:} Conceptualization; Formal analysis;
Investigation; Methodology; Software; Supervision; Validation;
Visualization; Writing - original draft; and Writing - review \&
editing. \textbf{Sindre Stenen Blakseth:} Conceptualization; Formal
analysis; Investigation; Methodology; Software; Validation;
Visualization; Writing - original draft; and Writing - review \&
editing. \textbf{André Massing:} Funding acquisition; Methodology;
Supervision; Writing - review \& editing. \textbf{Petter Nekså:}
Funding acquisition; Supervision; Writing - review \&
editing. \textbf{Magnus Aa.\ Gjennestad:} Conceptualization; Formal
analysis; Investigation; Methodology; Project administration;
Software; Supervision; Validation; Visualization; Writing - original
draft; and Writing - review \& editing.

\bibliographystyle{elsarticle-num}

\end{document}

%% file: preamble.tex
\usepackage{amsmath,amssymb,mathtools}
\usepackage[colorinlistoftodos]{todonotes}
\presetkeys{todonotes}{inline}{}
\usepackage{placeins} 
\usepackage{booktabs}
\usepackage{xspace}
\usepackage{fullpage}

\usepackage[utf8]{inputenc}
\usepackage[hidelinks]{hyperref}
\usepackage{color}
\usepackage[version=4]{mhchem}
\usepackage{booktabs}

\usepackage{subcaption}
\usepackage{listings,xspace}
\usepackage{pdfpages}
\usepackage{verbatim}
\usepackage{url}
\usepackage{enumerate}
\usepackage{physics}
\usepackage{multirow}

\usepackage{tikz}
\usetikzlibrary{circuits.ee.IEC, positioning}
\graphicspath{{./gfx/}}

\usepackage[per-mode=symbol]{siunitx}
\DeclareSIUnit\atm{atm}

\usepackage{cleveref}
\crefname{equation}{Eq.}{Eqs.}
\crefname{section}{Sec.}{Secs.}
\crefname{table}{Tab.}{Tabs.}
\crefname{figure}{Fig.}{Figs.}
\crefname{subfigure}{Fig.}{Figs.}

\usepackage{caption}
\newcommand{\Tw}{T_{\text{w}}}
\newcommand{\Tc}{T_{\text{c}}}
\newcommand{\Tm}{T_{\text{m}}}

\newcommand{\xh}{x_{\text{h}}}
\newcommand{\xc}{x_{\text{c}}}

\newcommand{\lhto}{\ce{LH2}\xspace}

\newcommand{\gas}{{\text{gas}}}
\newcommand{\solid}{{\text{sol}}}
\newcommand{\rad}{{\text{rad}}}
\newcommand{\cond}{{\text{cond}}}
\newcommand{\conv}{{\text{conv}}}

\newcommand{\eff}{{\text{eff}}}
\newcommand{\keff}{{\ensuremath k_{\text{eff}}}}
\newcommand{\supp}{{\text{supp}}}
\newcommand{\ins}{{\text{ins}}}

\newcommand{\coldspot}{{\text{cs}}}

\newcommand{\pc}[1]{\ensuremath\left\{ #1 \right\}}
\newcommand{\pp}[1]{\ensuremath\left( #1 \right)}

\newcommand{\vap}{{\text{vap}}}
\newcommand{\htc}{{h}}

\newcommand{\Kn}{{\text{Kn}}}

\newcommand{\amb}{{\text{amb}}}

\newcommand{\bor}{\text{BOR}}

\renewcommand{\vec}[1]{\ensuremath{\boldsymbol{ #1 }}}

%% file: manuscript-arxiv.bbl
\begin{thebibliography}{10}
\expandafter\ifx\csname url\endcsname\relax
  \def\url#1{\texttt{#1}}\fi
\expandafter\ifx\csname urlprefix\endcsname\relax\def\urlprefix{URL }\fi
\expandafter\ifx\csname href\endcsname\relax
  \def\href#1#2{#2} \def\path#1{#1}\fi

\bibitem{stolten2016hydrogen}
D.~Stolten, B.~Emonts, \href{http://dx.doi.org/10.1002/9783527674268}{Hydrogen
  Science and Engineering: Materials, Processes, Systems and Technology},
  Wiley, 2016.
\newblock \href {https://doi.org/10.1002/9783527674268}
  {\path{doi:10.1002/9783527674268}}.
\newline\urlprefix\url{http://dx.doi.org/10.1002/9783527674268}

\bibitem{Yin2024}
L.~Yin, H.~Yang, Y.~Ju,
  \href{http://dx.doi.org/10.1016/j.ijhydene.2024.01.093}{Review on the key
  technologies and future development of insulation structure for liquid
  hydrogen storage tanks}, International Journal of Hydrogen Energy 57 (2024)
  1302--1315.
\newblock \href {https://doi.org/10.1016/j.ijhydene.2024.01.093}
  {\path{doi:10.1016/j.ijhydene.2024.01.093}}.
\newline\urlprefix\url{http://dx.doi.org/10.1016/j.ijhydene.2024.01.093}

\bibitem{Ma2024}
N.~Ma, W.~Zhao, W.~Wang, X.~Li, H.~Zhou,
  \href{http://dx.doi.org/10.1016/j.ijhydene.2023.09.021}{Large scale of green
  hydrogen storage: Opportunities and challenges}, International Journal of
  Hydrogen Energy 50 (2024) 379--396.
\newblock \href {https://doi.org/10.1016/j.ijhydene.2023.09.021}
  {\path{doi:10.1016/j.ijhydene.2023.09.021}}.
\newline\urlprefix\url{http://dx.doi.org/10.1016/j.ijhydene.2023.09.021}

\bibitem{Muthukumar2023}
P.~Muthukumar, A.~Kumar, M.~Afzal, S.~Bhogilla, P.~Sharma, A.~Parida, S.~Jana,
  E.~A. Kumar, R.~K. Pai, I.~Jain,
  \href{http://dx.doi.org/10.1016/j.ijhydene.2023.04.304}{Review on large-scale
  hydrogen storage systems for better sustainability}, International Journal of
  Hydrogen Energy 48~(85) (2023) 33223--33259.
\newblock \href {https://doi.org/10.1016/j.ijhydene.2023.04.304}
  {\path{doi:10.1016/j.ijhydene.2023.04.304}}.
\newline\urlprefix\url{http://dx.doi.org/10.1016/j.ijhydene.2023.04.304}

\bibitem{karayel2023comprehensive}
G.~K. Karayel, N.~Javani, I.~Dincer, A comprehensive assessment of energy
  storage options for green hydrogen, Energy Conversion and Management 291
  (2023) 117311.

\bibitem{Andersson2019}
J.~Andersson, S.~Gr\"{o}nkvist, Large-scale storage of hydrogen, International
  Journal of Hydrogen Energy 44~(23) (2019) 11901--11919.
\newblock \href {https://doi.org/10.1016/j.ijhydene.2019.03.063}
  {\path{doi:10.1016/j.ijhydene.2019.03.063}}.

\bibitem{preuster2017liquid}
P.~Preuster, C.~Papp, P.~Wasserscheid, Liquid organic hydrogen carriers
  {(LOHCs)}: {T}oward a hydrogen-free hydrogen economy, Accounts of Chemical
  Research 50~(1) (2017) 74--85.
\newblock \href {https://doi.org/10.1021/acs.accounts.6b00474}
  {\path{doi:10.1021/acs.accounts.6b00474}}.

\bibitem{zheng2021current}
J.~Zheng, H.~Zhou, C.-G. Wang, E.~Ye, J.~W. Xu, X.~J. Loh, Z.~Li, Current
  research progress and perspectives on liquid hydrogen rich molecules in
  sustainable hydrogen storage, Energy Storage Materials 35 (2021) 695--722.

\bibitem{park202220}
J.~Park, J.~Ha, R.~Muhammad, H.~K. Lee, R.~Balderas-Xicohtencatl, Y.~Cheng,
  A.~J. Ramirez-Cuesta, B.~Streppel, M.~Hirscher, H.~R. Moon, H.~Oh, 20 {K}
  {H}$_2$ physisorption on metal-organic frameworks with enhanced dormancy
  compared to liquid hydrogen storage, ACS Applied Energy Materials 6~(18)
  (2022).
\newblock \href {https://doi.org/10.1021/acsaem.2c01907}
  {\path{doi:10.1021/acsaem.2c01907}}.

\bibitem{krasae2010development}
S.~Krasae-in, J.~H. Stang, P.~Neksa, Development of large-scale hydrogen
  liquefaction processes from 1898 to 2009, International Journal of Hydrogen
  Energy 35~(10) (2010) 4524--4533.
\newblock \href {https://doi.org/10.1016/j.ijhydene.2010.02.109}
  {\path{doi:10.1016/j.ijhydene.2010.02.109}}.

\bibitem{al2022hydrogen}
S.~Z. Al~Ghafri, S.~Munro, U.~Cardella, T.~Funke, W.~Notardonato, J.~P.~M.
  Trusler, J.~Leachman, R.~Span, S.~Kamiya, G.~Pearce, A.~Swanger, E.~D.
  Rodriguez, P.~Bajada, F.~Jiao, K.~Peng, A.~Siahvashi, M.~L. Johns, E.~F. May,
  Hydrogen liquefaction: a review of the fundamental physics, engineering
  practice and future opportunities, Energy \& Environmental Science 15~(7)
  (2022) 2690--2731.
\newblock \href {https://doi.org/10.1039/d2ee00099g}
  {\path{doi:10.1039/d2ee00099g}}.

\bibitem{ustolin2022extensive}
F.~Ustolin, A.~Campari, R.~Taccani, An extensive review of liquid hydrogen in
  transportation with focus on the maritime sector, Journal of Marine Science
  and Engineering 10~(9) (2022) 1222.
\newblock \href {https://doi.org/10.3390/jmse10091222}
  {\path{doi:10.3390/jmse10091222}}.

\bibitem{berstad2009comparison}
D.~O. Berstad, J.~H. Stang, P.~Neks{\aa}, Comparison criteria for large-scale
  hydrogen liquefaction processes, International Journal of Hydrogen Energy
  34~(3) (2009) 1560--1568.
\newblock \href {https://doi.org/10.1016/j.ijhydene.2008.11.058}
  {\path{doi:10.1016/j.ijhydene.2008.11.058}}.

\bibitem{Cardella2017b}
U.~Cardella, L.~Decker, J.~Sundberg, H.~Klein, Process optimization for
  large-scale hydrogen liquefaction, International Journal of Hydrogen Energy
  42~(17) (2017) 12339--12354.
\newblock \href {https://doi.org/10.1016/j.ijhydene.2017.03.167}
  {\path{doi:10.1016/j.ijhydene.2017.03.167}}.

\bibitem{Wilhelmsen2018}
{\O}.~Wilhelmsen, D.~Berstad, A.~Aasen, P.~Neks{\aa}, G.~Skaugen, Reducing the
  exergy destruction in the cryogenic heat exchangers of hydrogen liquefaction
  processes, International Journal of Hydrogen Energy 43~(10) (2018)
  5033--5047.
\newblock \href {https://doi.org/10.1016/j.ijhydene.2018.01.094}
  {\path{doi:10.1016/j.ijhydene.2018.01.094}}.

\bibitem{Leachman2009}
J.~W. Leachman, R.~T. Jacobsen, S.~G. Penoncello, E.~W. Lemmon, {Fundamental
  equations of state for parahydrogen, normal hydrogen, and orthohydrogen},
  Journal of Physical and Chemical Reference Data 38~(3) (2009) 721--748.
\newblock \href {https://doi.org/10.1063/1.3160306}
  {\path{doi:10.1063/1.3160306}}.

\bibitem{Span2000Nitrogen}
R.~Span, E.~W. Lemmon, R.~T. Jacobsen, W.~Wagner, A.~Yokozeki,
  \href{http://dx.doi.org/10.1063/1.1349047}{A reference equation of state for
  the thermodynamic properties of nitrogen for temperatures from 63.151 to 1000
  k and pressures to 2200 mpa}, Journal of Physical and Chemical Reference Data
  29~(6) (2000) 1361--1433.
\newblock \href {https://doi.org/10.1063/1.1349047}
  {\path{doi:10.1063/1.1349047}}.
\newline\urlprefix\url{http://dx.doi.org/10.1063/1.1349047}

\bibitem{Setzmann1991Methane}
U.~Setzmann, W.~Wagner, \href{http://dx.doi.org/10.1063/1.555898}{A new
  equation of state and tables of thermodynamic properties for methane covering
  the range from the melting line to 625 k at pressures up to 1000 mpa},
  Journal of Physical and Chemical Reference Data 20~(6) (1991) 1061--1155.
\newblock \href {https://doi.org/10.1063/1.555898}
  {\path{doi:10.1063/1.555898}}.
\newline\urlprefix\url{http://dx.doi.org/10.1063/1.555898}

\bibitem{bell2014pure}
I.~H. Bell, J.~Wronski, S.~Quoilin, V.~Lemort, Pure and pseudo-pure fluid
  thermophysical property evaluation and the open-source thermophysical
  property library {CoolProp}, Industrial \& Engineering Chemistry Research
  53~(6) (2014) 2498--2508.
\newblock \href {https://doi.org/10.1021/ie4033999}
  {\path{doi:10.1021/ie4033999}}.

\bibitem{Ratnakar2021}
R.~R. Ratnakar, N.~Gupta, K.~Zhang, C.~van Doorne, J.~Fesmire, B.~Dindoruk,
  V.~Balakotaiah, Hydrogen supply chain and challenges in large-scale {LH2}
  storage and transportation, International Journal of Hydrogen Energy (2021).
\newblock \href {https://doi.org/10.1016/j.ijhydene.2021.05.025}
  {\path{doi:10.1016/j.ijhydene.2021.05.025}}.

\bibitem{Adler2024b}
E.~J. Adler, J.~R. R.~A. Martins, Liquid hydrogen tank boil-off model for
  design and optimization, Journal of Thermophysics and Heat Transfer (November
  2024).

\bibitem{joseph2017effect}
J.~Joseph, G.~Agrawal, D.~K. Agarwal, J.~Pisharady, S.~Sunil~Kumar, Effect of
  insulation thickness on pressure evolution and thermal stratification in a
  cryogenic tank, Applied Thermal Engineering 111 (2017) 1629--1639.
\newblock \href {https://doi.org/10.1016/j.applthermaleng.2016.07.015}
  {\path{doi:10.1016/j.applthermaleng.2016.07.015}}.

\bibitem{al2022modelling}
S.~Z.~S. Al~Ghafri, A.~Swanger, V.~Jusko, A.~Siahvashi, F.~Perez, M.~L. Johns,
  E.~F. May, Modelling of liquid hydrogen boil-off, Energies 15~(3) (2022)
  1149.
\newblock \href {https://doi.org/10.3390/en15031149}
  {\path{doi:10.3390/en15031149}}.

\bibitem{majumdar2008numerical}
A.~Majumdar, T.~Steadman, J.~Maroney, J.~Sass, J.~Fesmire, Numerical modeling
  of propellant boil-off in a cryogenic storage tank, in: AIP Conference
  Proceedings, Vol. 985, American Institute of Physics, 2008, pp. 1507--1514.
\newblock \href {https://doi.org/10.1063/1.2908513}
  {\path{doi:10.1063/1.2908513}}.

\bibitem{Barsi2008}
S.~Barsi, M.~Kassemi, Numerical and experimental comparisons of the
  self-pressurization behavior of an {LH2} tank in normal gravity, Cryogenics
  48 (2008) 122--129.
\newblock \href {https://doi.org/10.1016/j.cryogenics.2008.01.003}
  {\path{doi:10.1016/j.cryogenics.2008.01.003}}.

\bibitem{barsi2013investigation}
S.~Barsi, M.~Kassemi, Investigation of tank pressurization and pressure
  control--{Part} {II}: numerical modeling, Journal of Thermal Science and
  Engineering Applications 5~(4) (2013) 041006.
\newblock \href {https://doi.org/10.1115/1.4023892}
  {\path{doi:10.1115/1.4023892}}.

\bibitem{Matveev2023}
K.~I. Matveev, J.~W. Leachman,
  \href{http://dx.doi.org/10.3390/hydrogen4030030}{The effect of liquid
  hydrogen tank size on self-pressurization and constant-pressure venting},
  Hydrogen 4~(3) (2023) 444--455.
\newblock \href {https://doi.org/10.3390/hydrogen4030030}
  {\path{doi:10.3390/hydrogen4030030}}.
\newline\urlprefix\url{http://dx.doi.org/10.3390/hydrogen4030030}

\bibitem{mendez2021enabling}
E.~Mendez~Ramos, Enabling conceptual design and analysis of cryogenic in-space
  vehicles through the development of an extensible boil-off model (2021).

\bibitem{ratnakar2023effective}
R.~R. Ratnakar, Z.~Sun, V.~Balakotaiah, Effective thermal conductivity of
  insulation materials for cryogenic {LH}$_2$ storage tanks: {A} review,
  International Journal of Hydrogen Energy 48~(21) (2023) 7770--7793.
\newblock \href {https://doi.org/10.1016/j.ijhydene.2022.11.130}
  {\path{doi:10.1016/j.ijhydene.2022.11.130}}.

\bibitem{Babac2009}
G.~Babac, A.~Sisman, T.~Cimen, {Two-dimensional thermal analysis of liquid
  hydrogen tank insulation}, International Journal of Hydrogen Energy 34 (2009)
  6357--6363.
\newblock \href {https://doi.org/10.1016/j.ijhydene.2009.05.052}
  {\path{doi:10.1016/j.ijhydene.2009.05.052}}.

\bibitem{zheng2019thermodynamic}
J.~Zheng, L.~Chen, J.~Wang, X.~Xi, H.~Zhu, Y.~Zhou, J.~Wang, Thermodynamic
  analysis and comparison of four insulation schemes for liquid hydrogen
  storage tank, Energy Conversion and Management 186 (2019) 526--534.
\newblock \href {https://doi.org/10.1016/j.enconman.2019.02.073}
  {\path{doi:10.1016/j.enconman.2019.02.073}}.

\bibitem{Zheng2019}
J.~Zheng, L.~Chen, C.~Cui, Y.~Zhou, J.~Wang, {Calculation and position
  optimization on vapor cooled shield for liquid hydrogen storage}, in: IOP
  Conference Series: Materials Science and Engineering, Vol. 502, 2019, p.
  012080.
\newblock \href {https://doi.org/10.1088/1757-899X/502/1/012080}
  {\path{doi:10.1088/1757-899X/502/1/012080}}.

\bibitem{Jiang2021}
W.~Jiang, P.~Sun, P.~Li, Z.~Zuo, Y.~Huang,
  \href{http://dx.doi.org/10.1016/j.energy.2021.120859}{Transient thermal
  behavior of multi-layer insulation coupled with vapor cooled shield used for
  liquid hydrogen storage tank}, Energy 231 (2021) 120859.
\newblock \href {https://doi.org/10.1016/j.energy.2021.120859}
  {\path{doi:10.1016/j.energy.2021.120859}}.
\newline\urlprefix\url{http://dx.doi.org/10.1016/j.energy.2021.120859}

\bibitem{liu2016thermal}
Z.~Liu, Y.~Li, F.~Xie, K.~Zhou, Thermal performance of foam/mli for cryogenic
  liquid hydrogen tank during the ascent and on orbit period, Applied Thermal
  Engineering 98 (2016) 430--439.

\bibitem{Hofmann2006}
A.~Hofmann, {The thermal conductivity of cryogenic insulation materials and its
  temperature dependence}, Cryogenics 46 (2006) 815--824.
\newblock \href {https://doi.org/10.1016/j.cryogenics.2006.08.001}
  {\path{doi:10.1016/j.cryogenics.2006.08.001}}.

\bibitem{Wang2020}
Z.~Wang, Y.~Wang, S.~Afshan, J.~Hjalmarsson, A review of metallic tanks for h2
  storage with a view to application in future green shipping, International
  Journal of Hydrogen Energy (2020).

\bibitem{tapeinos2016design}
I.~G. Tapeinos, S.~Koussios, R.~M. Groves, Design and analysis of a multi-cell
  subscale tank for liquid hydrogen storage, International Journal of Hydrogen
  Energy 41~(5) (2016) 3676--3688.
\newblock \href {https://doi.org/10.1016/j.ijhydene.2015.10.104}
  {\path{doi:10.1016/j.ijhydene.2015.10.104}}.

\bibitem{tapeinos2019evaluationI}
I.~G. Tapeinos, D.~S. Zarouchas, O.~K. Bergsma, S.~Koussios, R.~Benedictus,
  Evaluation of the mechanical performance of a composite multi-cell tank for
  cryogenic storage: {P}art {I} - {T}ank pressure window based on progressive
  failure analysis, International Journal of Hydrogen Energy 44~(7) (2019)
  3917--3930.
\newblock \href {https://doi.org/10.1016/j.ijhydene.2018.12.118}
  {\path{doi:10.1016/j.ijhydene.2018.12.118}}.

\bibitem{tapeinos2019evaluationII}
I.~G. Tapeinos, A.~Rajabzadeh, D.~S. Zarouchas, M.~Stief, R.~M. Groves,
  S.~Koussios, R.~Benedictus, Evaluation of the mechanical performance of a
  composite multi-cell tank for cryogenic storage: {P}art {II} --
  {E}xperimental assessment, International Journal of Hydrogen Energy 44~(7)
  (2019) 3931--3943.
\newblock \href {https://doi.org/10.1016/j.ijhydene.2018.12.063}
  {\path{doi:10.1016/j.ijhydene.2018.12.063}}.

\bibitem{johnson2022analysis}
W.~Johnson, R.~Balasubramaniam, R.~Grotenrath, Analysis of heat transfer from a
  local heat source at cryogenic temperatures, in: IOP Conference Series:
  Materials Science and Engineering, Vol. 1240, 2022, p. 012013.
\newblock \href {https://doi.org/10.1088/1757-899X/1240/1/012013}
  {\path{doi:10.1088/1757-899X/1240/1/012013}}.

\bibitem{Mantzaroudis2023}
V.~K. Mantzaroudis, E.~E. Theotokoglou, Computational analysis of liquid
  hydrogen storage tanks for aircraft applications, Materials 16~(6) (2023).
\newblock \href {https://doi.org/10.3390/ma16062245}
  {\path{doi:10.3390/ma16062245}}.

\bibitem{Abe1998}
A.~Abe, M.~Nakamura, I.~Sato, H.~Uetani, T.~Fujitani, {Studies of the
  large-scale sea transportation of liquid hydrogen}, International Journal of
  Hydrogen Energy 23 (1998) 115--121.
\newblock \href {https://doi.org/10.1016/s0360-3199(97)00032-3}
  {\path{doi:10.1016/s0360-3199(97)00032-3}}.

\bibitem{Scurlock2017}
R.~Scurlock, {Problems of bulk storage and shipping of liquid hydrogen in
  volumes of 10,000 to 100,000 cubic meters at one bar pressure}, in:
  {Cryogenics 2017. Proceedings of the 14th IIR International Conference},
  {International Institute of Refrigeration}, 2017, pp. 140--144.
\newblock \href {https://doi.org/10.18462/iir.cryo.2017.0001}
  {\path{doi:10.18462/iir.cryo.2017.0001}}.

\bibitem{alkhaledi2022hydrogen}
A.~N. Alkhaledi, S.~Sampath, P.~Pilidis, A hydrogen fuelled {LH2} tanker ship
  design, Ships and Offshore Structures 17~(7) (2022) 1555--1564.
\newblock \href {https://doi.org/10.1080/17445302.2021.1935626}
  {\path{doi:10.1080/17445302.2021.1935626}}.

\bibitem{Cristea2020}
V.~Cristea, Storage tank concepts for liquefied hydrogen, Master's thesis,
  {Norwegian University of Science and Technology} (2020).

\bibitem{barron2017cryogenic}
R.~F. Barron, G.~F. Nellis, Cryogenic heat transfer, CRC press, 2017.

\bibitem{NISTCryogenicsWebpage}
NIST,
  \href{https://trc.nist.gov/cryogenics/materials/materialproperties.htm}{Cryogenic
  technology resources - {M}aterial properties} (2021).
\newline\urlprefix\url{https://trc.nist.gov/cryogenics/materials/materialproperties.htm}

\bibitem{incropera_ed6}
F.~P. Incropera, D.~P. DeWitt, T.~L. Bergman, A.~S. Lavine, Fundamentals of
  heat and mass transfer, 6th Edition, Wiley, 2007.

\bibitem{Lienhard2019}
J.~H. Lienhard~IV, J.~H. Lienhard~V, A Heat Transfer Textbook, 5th Edition,
  Phlogiston Press, 2019.

\bibitem{ogunsola2015application}
O.~T. Ogunsola, L.~Song, Application of a simplified thermal network model for
  real-time thermal load estimation, Energy and buildings 96 (2015) 309--318.
\newblock \href {https://doi.org/10.1016/j.enbuild.2015.03.044}
  {\path{doi:10.1016/j.enbuild.2015.03.044}}.

\bibitem{laug}
E.~Anderson, Z.~Bai, C.~Bischof, S.~Blackford, J.~Demmel, J.~Dongarra,
  J.~Du~Croz, A.~Greenbaum, S.~Hammarling, A.~McKenney, D.~Sorensen, {LAPACK}
  Users' Guide, 3rd Edition, Society for Industrial and Applied Mathematics,
  Philadelphia, PA, 1999.

\bibitem{harris2020array}
C.~R. Harris, K.~J. Millman, S.~J. van~der Walt, R.~Gommers, P.~Virtanen,
  D.~Cournapeau, E.~Wieser, J.~Taylor, S.~Berg, N.~J. Smith, R.~Kern, M.~Picus,
  S.~Hoyer, M.~H. van Kerkwijk, M.~Brett, A.~Haldane, J.~F. del R{\'{i}}o,
  M.~Wiebe, P.~Peterson, P.~G{\'{e}}rard-Marchant, K.~Sheppard, T.~Reddy,
  W.~Weckesser, H.~Abbasi, C.~Gohlke, T.~E. Oliphant,
  \href{https://doi.org/10.1038/s41586-020-2649-2}{Array programming with
  {NumPy}}, Nature 585~(7825) (2020) 357--362.
\newblock \href {https://doi.org/10.1038/s41586-020-2649-2}
  {\path{doi:10.1038/s41586-020-2649-2}}.
\newline\urlprefix\url{https://doi.org/10.1038/s41586-020-2649-2}

\bibitem{schoberl1997netgen}
J.~Sch{\"o}berl, {NETGEN} {A}n advancing front {2D/3D-mesh} generator based on
  abstract rules, Computing and Visualization in Science 1 (1997) 41--52.
\newblock \href {https://doi.org/10.1007/s007910050004}
  {\path{doi:10.1007/s007910050004}}.

\bibitem{schoberl2014c++}
J.~Sch{\"o}berl, {C++ 11} implementation of finite elements in {NGSolve}, Tech.
  Rep. 30/2014, Institute for analysis and scientific computing -- Vienna
  University of Technology (2014).

\bibitem{simon1992copper}
N.~J. Simon, E.~S. Drexler, R.~P. Reed, Properties of copper and copper alloys
  at cryogenic temperatures. final report, Tech. rep. (2 1992).
\newblock \href {https://doi.org/10.2172/5340308} {\path{doi:10.2172/5340308}}.

\bibitem{woodcraft2005aluminium}
A.~L. Woodcraft, Recommended values for the thermal conductivity of aluminium
  of different purities in the cryogenic to room temperature range, and a
  comparison with copper, Cryogenics 45~(9) (2005) 626--636.
\newblock \href {https://doi.org/10.1016/j.cryogenics.2005.06.008}
  {\path{doi:10.1016/j.cryogenics.2005.06.008}}.

\bibitem{ratnakar2023short}
R.~R. Ratnakar, S.~Sharma, M.~Taghavi, V.~Balakotaiah, A short discussion on
  insulation strategies and design considerations for large-scale storage and
  transportation of liquid hydrogen, Energy Systems (2023) 1--13\href
  {https://doi.org/10.1007/s12667-023-00649-1}
  {\path{doi:10.1007/s12667-023-00649-1}}.

\bibitem{Fesmire2015}
J.~E. Fesmire, Standardization in cryogenic insulation systems testing and
  performance data, Physics Procedia 67 (2015) 1089--1097.
\newblock \href {https://doi.org/10.1016/j.phpro.2015.06.205}
  {\path{doi:10.1016/j.phpro.2015.06.205}}.

\bibitem{Skogan2022}
T.~Skogan, J.~Hveem, F.~Fløtten, J.~Anker,
  \href{https://search.patentstyret.no/Patent/20221390}{Liquified gas storage
  tank}, patent application no. {NO} 20221390. {Norwegian Industrial Property
  Office}, 2022-12-22.
\newline\urlprefix\url{https://search.patentstyret.no/Patent/20221390}

\bibitem{SanMarchi2012}
C.~{San Marchi}, B.~P. Somerday, Technical reference for hydrogen compatibility
  of materials, Tech. rep., Sandia National Laboratories, {SAND2012-7321}
  (2012).

\bibitem{Krenn2012}
A.~G. Krenn, Diagnosis of a poorly performing liquid hydrogen bulk storage
  sphere, in: AIP conference proceedings, Vol. 1434, American Institute of
  Physics, 2012, pp. 376--383.
\newblock \href {https://doi.org/10.1063/1.4706942}
  {\path{doi:10.1063/1.4706942}}.

\bibitem{Bostock2019}
T.~D. Bostock, R.~G. Scurlock, Low-Loss Storage and Handling of Cryogenic
  Liquids, Springer, 2019.

\bibitem{berstad2022liquid}
D.~Berstad, S.~Gardarsdottir, S.~Roussanaly, M.~Voldsund, Y.~Ishimoto,
  P.~Neks{\aa}, Liquid hydrogen as prospective energy carrier: {A} brief review
  and discussion of underlying assumptions applied in value chain analysis,
  Renewable and Sustainable Energy Reviews 154 (2022) 111772.
\newblock \href {https://doi.org/10.1016/j.rser.2021.111772}
  {\path{doi:10.1016/j.rser.2021.111772}}.

\bibitem{Sass2010}
J.~Sass, W.~S. Cyr, T.~Barrett, R.~Baumgartner, J.~Lott, J.~Fesmire, Glass
  bubbles insulation for liquid hydrogen storage tanks, in: {AIP Conference
  Proceedings}, Vol. 1218, American Institute of Physics, 2010, pp. 772--779.
\newblock \href {https://doi.org/10.1063/1.3422430}
  {\path{doi:10.1063/1.3422430}}.

\bibitem{Sass2008}
J.~Sass, J.~Fesmire, Z.~Nagy, S.~Sojourner, D.~Morris, S.~Augustynowicz,
  Thermal performance comparison of glass microsphere and perlite insulation
  systems for liquid hydrogen storage tanks, in: AIP conference proceedings,
  Vol. 985, American Institute of Physics, 2008, pp. 1375--1382.
\newblock \href {https://doi.org/10.1063/1.2908497}
  {\path{doi:10.1063/1.2908497}}.

\end{thebibliography}
